\documentclass[journal]{IEEEtran}
\usepackage{graphicx}
\usepackage{booktabs}
\usepackage{subfigure}
\usepackage{fontawesome}
\usepackage{overpic}
\usepackage{amsmath}
\usepackage{amssymb}
\usepackage{multirow}
\usepackage{makecell}
\usepackage{array}
\usepackage{stfloats}
\usepackage{listings}
\usepackage{algorithm}
\usepackage{color, soul}
\usepackage{bbding}
\usepackage[table]{xcolor}
\usepackage{graphicx}
\usepackage{algpseudocode}
\usepackage{amsmath} 
\usepackage{amssymb} 
\usepackage{hyperref}
\usepackage{cleveref} 
\usepackage{bm}
\usepackage{multicol}
\usepackage{cite}

\usepackage{tabularx}
\renewcommand{\algorithmicensure}{\textbf{Output:}} 
\renewcommand{\algorithmicrequire}{\textbf{Input:}}

\hypersetup{
  colorlinks=true,
  linkcolor=blue,
  urlcolor=blue,
  citecolor=blue
}

\begin{document}

\title{MSFMamba: Multi-Scale Feature Fusion State Space Model for Multi-Source Remote Sensing Image Classification}

\author{
Feng Gao, \textit{Member, IEEE}, 
Xuepeng Jin, \textit{Student Member, IEEE},
Xiaowei Zhou, \\
Junyu Dong, \textit{Member, IEEE},
Qian Du \textit{Fellow, IEEE}
\thanks{This work was supported in part by the National Science and Technology Major Project under Grant 2022ZD0117202, in part by the Natural Science Foundation of Qingdao under Grant 23-2-1-222-ZYYD-JCH, and in part by the Postdoctoral Fellowship Program of CPSF under Grant GZC20241614. (\textit{Corresponding author: Xiaowei Zhou.})

Feng Gao, Xuepeng Jin, Xiaowei Zhou, and Junyu Dong are with the School of Computer Science and Technology, Ocean University of China, Qingdao 266100, China. (email: gaofeng@ouc.edu.cn, 2231693339@qq.com, zhouxiaowei@ouc.edu.cn, dongjunyu@ouc.edu.cn)

Qian Du is with the Department of Electrical and Computer Engineering, Mississippi State University, Starkville, MS 39762 USA. (email: du@ece.msstate.edu)}}

\markboth{IEEE Transactions on Geoscience and Remote Sensing}{Shell}

\maketitle

\begin{abstract}
In the field of multi-source remote sensing image classification, remarkable progress has been made by using {Convolutional Neural Network (CNN)} and Transformer.
While CNNs are constrained by their local receptive fields, Transformers mitigate this issue with their global attention mechanism. However, Transformers come with the trade-off of higher computational complexity.
Recently, Mamba-based methods built upon the State Space Model (SSM) have shown great potential for long-range dependency modeling with linear complexity, but they have rarely been explored for multi-source remote sensing image classification tasks. 
To address this issue, we propose the Multi-Scale Feature Fusion Mamba (MSFMamba) network, a novel framework designed for the joint classification of hyperspectral image (HSI) and Light Detection and Ranging (LiDAR)/Synthetic Aperture Radar (SAR) data. The MSFMamba network is composed of three key components: the Multi-Scale Spatial Mamba (MSpa-Mamba) block, the Spectral Mamba (Spe-Mamba) block, and the Fusion Mamba (Fus-Mamba) block.
The MSpa-Mamba block employs a multi-scale strategy to reduce computational cost and alleviate feature redundancy in multiple scanning routes, ensuring efficient spatial feature modeling. The Spe-Mamba block focuses on spectral feature extraction, addressing the unique challenges of HSI data representation. Finally, the Fus-Mamba block bridges the heterogeneous gap between HSI and LiDAR/SAR data by extending the original Mamba architecture to accommodate dual inputs, enhancing cross-modal feature interactions and enabling seamless data fusion.
Together, these components enable MSFMamba to effectively tackle the challenges of multi-source data classification, delivering improved performance with optimized computational efficiency.
Comprehensive experiments on four real-world multi-source remote sensing datasets (Berlin, Augsburg, Houston2018, and Houston2013) demonstrate the superiority of MSFMamba outperforms several state-of-the-art methods and achieves overall accuracies of 76.92\%, 91.38\%, 92.38\%, and 92.86\%, respectively. The source codes of MSFMamba will be publicly available at \url{https://github.com/oucailab/MSFMamba}.

\end{abstract}


\IEEEpeerreviewmaketitle

\begin{figure}[t]
    \centering
    \includegraphics[width=3.2in]{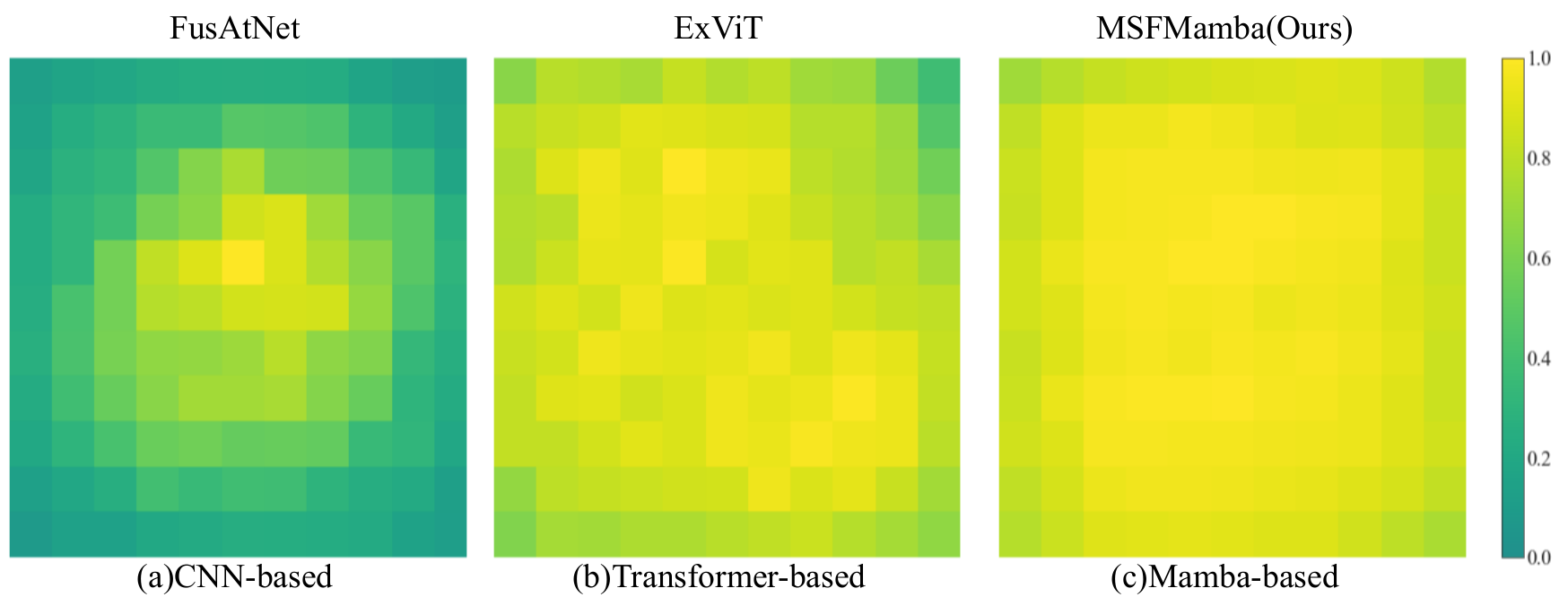}
    \caption{The effective receptive field of three models: (a) the CNN-based FusAtNet, (b) the Transformer-based ExViT, and (c) the proposed MSFMamba. A broader distribution of bright areas indicates larger effective receptive filed. The proposed MSFMamba achieves the largest effective receptive field.}
    \label{erf}
\end{figure}

\vspace{11pt}
\section{Introduction}

\IEEEPARstart{W}ITH the rapid development of remote sensing technology and sensor platforms, {multi-source remote sensing data from various platforms such as satellites, aircraft, and drones have become increasingly abundant \cite{zhmm17jars, weixin_join1, weixin_join2}.}These data have been widely applied in land use and land cover classification \cite{Latent_Reconstruction}, geological resource exploration \cite{mineral_exploration}, urban planning \cite{dynamic_Super-Pixel}, {disaster warning \cite{Terrain}, and agricultural monitoring \cite{agriculture} \cite{ganhan_yaogan_join1}}. Among these applications, land use and land cover classification is an important and fundamental task \cite{hzy24jstars} \cite{kk15jstars}.

Among these multi-source data captured by various sensors, HSI, LiDAR data, and SAR data are of great significance\cite{LiDAR_join1}. HSI can identify subtle differences among ground objects through its rich spectral information \cite{lz24tgrs,zx24grsl,hdf21grsm}. HSI classification has been widely studied, since the rich spectral information contained in HSI presents great opportunities and challenges. On the other hand, LiDAR data precisely measure the three-dimensional structure of terrain and objects with high spatial resolution and accuracy \cite{Range_Resolved} \cite{Terrestrial_Lidar}. SAR sensors perform imaging in all weather conditions and at all times, penetrating clouds and vegetation \cite{polarimetric_SAR,time-series_PolSAR,Multitemporal_SAR}. The fusion of HSI and LiDAR/SAR data can leverage the complementary advantages of multiple data sources and lead to improved ground object classification performance \cite{yj24tgrs, mxc24tgrs, gyh24tip, fusion_join2}. By combining the spectral information of HSI with the three-dimensional structural information of LiDAR or the all weather imaging capability of SAR, the limitations of a single data source can be alleviated and more reliable classification results can be achieved \cite{zmt16grsl} \cite{xzg18grsl}. Therefore, in this paper, we mainly focus on HSI and LiDAR/SAR data joint classification.

Because of diverse distributions and feature representations in multi-source images, the critical challenge for HSI and LiDAR/SAR data classification lies in how to bridge the heterogeneity gap. {Recently, tremendous efforts have been made for multi-source data classification \cite{xys24tgrs,lzy23grsl,Indian,DMSCA}.}The most common framework that demonstrated promising classification results employs CNN-based feature encoder in an end-to-end manner. Cross-guided attention \cite{dwq22tgrs}, cross-channel correlation \cite{wx22tgrs}, federated learning \cite{ldx24tgrs} and cross-scale mixing attention \cite{gyh23tgrs} are employed for multi-source data classification. CNN-based methods design complex attention mechanisms to capture useful features, but due to the inherent limitations of convolutional units, their receptive field range is limited, making it difficult to capture long-range dependencies. In addition, Transformer-based models have demonstrated outstanding long-range feature modeling capabilities in multi-source remote sensing data classification \cite{zly23jstars}. Spatial-spectral attention \cite{gyh22rs}, hierarchical attention \cite{xzx22tgrs} \cite{fyn24kbs}, scale-adaptive attention \cite{zmq23rs} are used for HSI and LiDAR/SAR data classification. Although the Transformer has a global receptive field, its attention mechanism has high computational complexity and is less efficient when handling long-range dependencies.

Recently, an improved SSM \cite{js23iclr} with a selective scanning mechanism, Mamba \cite{Mamba}, has emerged as an effective alternative for computer vision and remote sensing image interpretation. As shown in Fig. \ref{erf}, the Mamba-based method\cite{Mamba_tgrs1}\cite{Mamba_jstar1} performs better than CNN and Transformer-based methods in effective receptive field modeling.
By expanding the receptive field, the model is able to capture broader spatial dependencies, which is crucial for recognizing contextual objects and larger patterns in remote sensing data.
In contrast to Transformer-based models, the state-space sequence model reformulates the attention mechanism to scale linearly with sequence length, significantly reducing computational costs, especially for long sequences \cite{js24arxiv}. 

It is a non-trivial and challenging task to introduce the SSM into multi-source data classification, due to the following reasons: 1) \textbf{Feature redundancy in multiple scanning routes}. To address the challenge of image data nondirectionality, existing SSM-based models generally use a multi-scan strategy (e.g., forward, backward, horizontal, and vertical scan) to ensure that every part of the image can establish connections with other parts. However, the multi-scan strategy significantly increases feature redundancy, as similar patterns are repeatedly extracted across overlapping scanning routes. Multi-scale feature extraction provides a promising solution to this challenge by capturing spatial dependencies at different levels of granularity. By combining fine-grained local details with broader contextual information, the model can reduce redundancy while preserving the most informative features. Thus, exploring an effective multi-scale strategy is critical for improving the efficiency of SSM-based approaches.
2) \textbf{The Heterogeneous gap between multi-source data}. SSM's role in multi-source remote sensing data classification has not yet been fully explored, as SSM lacks a design similar to cross-attention. This prompts us to investigate how to use SSM to alleviate the heterogeneous gap between HSI and LiDAR/SAR data.


To address the above challenges, we propose \textbf{M}ulti-\textbf{S}cale Feature \textbf{F}usion \textbf{Mamba} (MSFMamba) network for the joint classification of HSI and LiDAR/SAR data. MSFMamba mainly comprises three parts: MSpa-Mamba block, Spe-Mamba block, and Fus-Mamba block. Specifically, to solve the feature redundancy in multiple scanning routes, the MSpa-Mamba block incorporates the multi-scale strategy to minimize the computational redundancy and alleviate the feature redundancy in SSM. In addition, Spe-Mamba is designed for spectral feature exploration, which is essential for HSI feature modeling. Moreover, to alleviate the heterogeneous gap between HSI and LiDAR/SAR data, we design Fus-Mamba block for multi-source feature fusion. The original Mamba is extended to accommodate dual inputs, and cross-modal feature interaction is enhanced. 

In summary, our main contributions can be summarized as follows:

\begin{itemize}

\item We propose MSFMamba, a simple yet effective sequential scanning model for HSI and LiDAR/SAR data interpretation, which investigates the use of SSMs for multi-source remote sensing image classification.

\item We design a MSpa-Mamba block that incorporates a multi-scale strategy to minimize computational burden, along with a Fus-Mamba block to enhance cross-modal feature interactions between multi-source data.

\item Extensive experiments on four public benchmark datasets demonstrate that our method outperforms state-of-the-art techniques in multi-source remote sensing image classification.

\end{itemize}

The remainder of this paper is organized as follows. Section II reviews related work. Section III introduces the preliminaries of the state space model and the detailed structure of MSFMamba. Experiments on four multi-source datasets are presented in Section IV. Conclusions are drawn, and future work is discussed in Section V.

\section{Related Works}
\subsection{Deep Learning-Based Multi-Source Image Classification}

Deep learning-based remote sensing image analysis has witnessed significant progress recently \cite{RSISCMDP, zgq23tgrs, zml23tbc, cdq22tcsvt}. Convolutional neural networks (CNNs) have been widely used in an end-to-end manner for multi-source feature fusion. Dong et al. \cite{dwq22tgrs} used self-attention and cross-guided attention to emphasize regions and channels of interest. Gao et al. \cite{gyh23tgrs} presented a cross-mixing attention network to extract multi-scale features for hyperspectral and multispectral image joint classification. Xu et al. \cite{TBCNN} introduced a dual-tunnel CNN architecture that separately extracts spectral-spatial features from HSI and LiDAR data. Zhang et al. \cite{PTop_CNN} proposed a model known as patch-to-patch CNN for multi-source data classification. They constructed a three-tower structure to compute multi-scale features. Mohla et al. \cite{FusAtNet} proposed Feature Fusion And exTraction Network (FusAtNet), which leverages HSI to generate an attention map via self-attention mechanism, which effectively highlights the intrinsic spectral features. In addition, cross-attention is employed to fuse the multi-source features. Fang et al. \cite{S$^2$ENet} presented a spatial–spectral enhancement module to promote multi-modal feature interaction. In addition, 

To enhance the long-range feature dependencies, many Transformer-based methods have been proposed for multi-source remote sensing image classification. Hu et al. \cite{hyx22igarss} proposed a simple yet effective parallel Transformer model. One Transformer acts as HSI feature encoder, and the other one is responsible for cross-modal feature interactions. Zhao et al. \cite{HCT} proposed a CNN and Transformer hybrid model for HSI and LiDAR data classification. Cross-token attention is used for cross-modal feature fusion. Yao et al. \cite{ExVit} proposed Extended Vision Transformer (ExViT), which modifies the traditional Vision Transformer to handle land use and land cover classification tasks. This framework processes multimodal remote sensing image data through parallel branches and introduces a cross-modality attention module to enhance information exchange between different modalities. Li et al. \cite{li2023mixing} introduced mixing attention and convolution network for multi-source feature fusion. Although these CNN and Transformer-based methods have gained performance improvement, there is still an urgent need for an efficient and effective scheme for heterogeneous representation. In this paper, we explore Mamba's potential for multi-source remote sensing feature fusion, and provide an effective paradigm to propel the progress of heterogeneous feature fusion.

\subsection{State Space Model} 

SSMs \cite{ga21nips}\cite{ga22iclr} have become practical components for constructing deep networks due to their cutting-edge performance in analyzing continuous long sequential data. Gu et al. \cite{ga22iclr} introduced a diagonal structure and combined it with a diagonal plus low-rank approach to construct structured SSM. Smith et al. \cite{js23iclr} improved the efficiency of SSM by introducing parallel scanning techniques. Mamba \cite{Mamba} incorporates data-dependent parameters to amend the linear time invariant characteristics of SSM-based models, and it has demonstrated excellent performance over Transformers on large-scale datasets.

Recently, inspired by pioneering SSMs, Mamba-like frameworks have been employed in computer vision and remote sensing image interpretation tasks, such as Visual State Space Model (VMamba) \cite{ly24vmamba}, remote sensing images semantic segmentation Mamba (RS$^3$Mamba) \cite{rs3mamba}, and Remote Sensing Mamba(RSMamba) \cite{cky24rsmamba}. RSMamba \cite{cky24rsmamba} employed dynamic multi-path activation mechanism to enhance the performance of remote sensing image classification. Zhang et al. \cite{zqf24arxiv} employed a unique encoder architecture, based on the Mamba design, to effectively extract semantic information from remote sensing images. Chen et al. \cite{chen24cm} used a Mamba-like feature encoder to learn global spatial contextual information for remote sensing image change detection. Li et al. \cite{lyp24tgrs} proposed a spectral Mamba block to extract the spectral features of HSI, and improve the HSI classification performance. These works show the potential for Mamba's extension into multi-source data fusion, but its particular application in the joint classification task of HSI and LiDAR/SAR data remains unexplored. 

\section{Methodology} 

In this section, we start by presenting the essential concepts of the state space model. Following that, we delve into a comprehensive description of our MSFMamba, covering its framework and module design in detail.

\begin{figure*}[t]  
\centering
\includegraphics [width=6in]{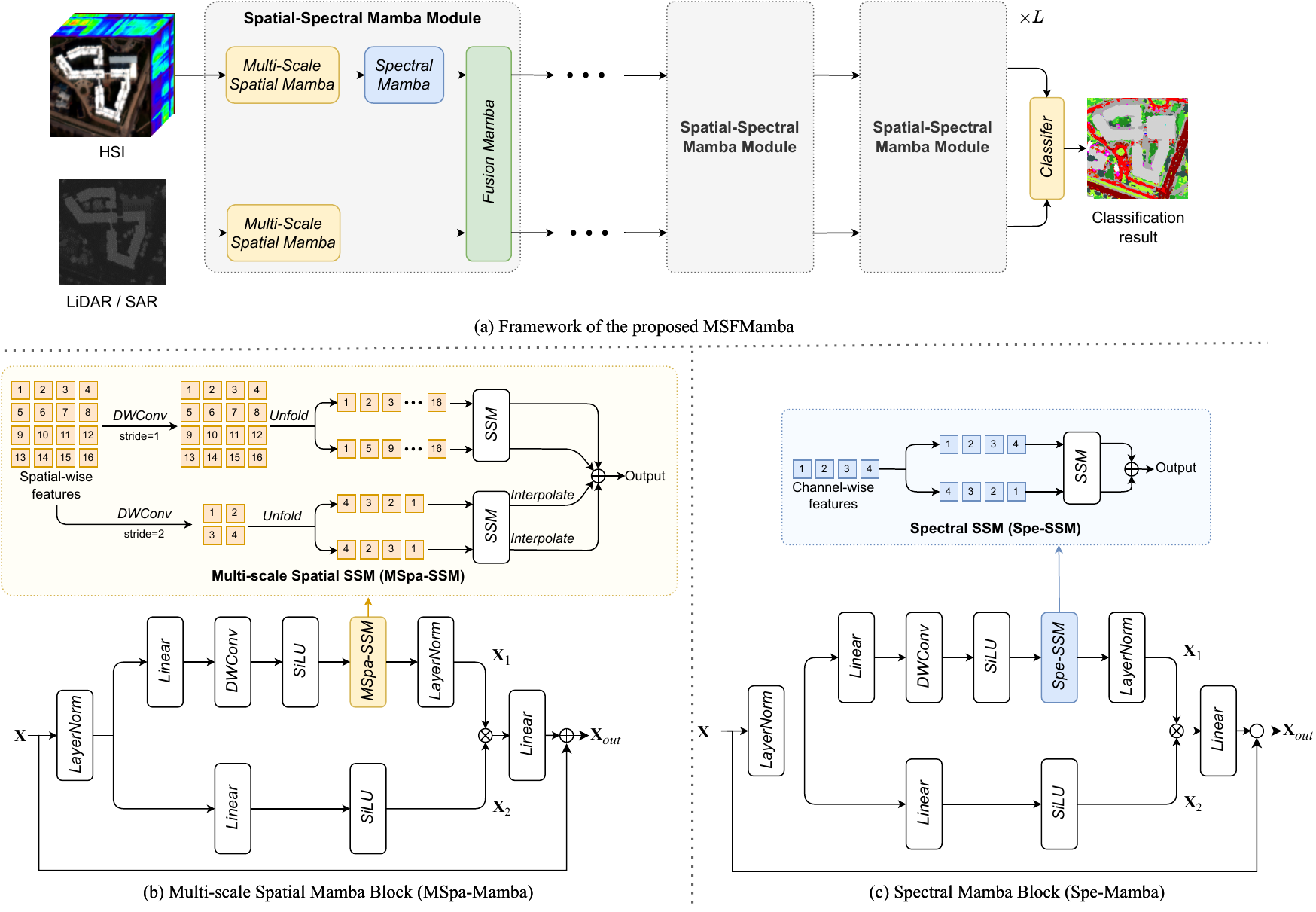}
\caption{Framework of the proposed Multi-Scale Feature Fusion Mamba (MSFMamba) for HSI and LiDAR/SAR data classification. (a) Details of the proposed MSFMamba. It contains $L$ spatial-spectral Mamba module to learn multi-source feature representations. Each spatial-spectral Mamba module is comprised of Multi-scale Spatial Mamba (MSpa-Mamba) block, Spectral Mamba (Spe-Mamba) block, and Fusion Mamba (Fus-Mamba) block. (b) Details of the MSpa-Mamba block, as well as the Multi-scale Spatial SSM (MSpa-SSM). (c) Details of the Spe-Mamba block, as well as the Spectral SSM (Spe-SSM).}
\label{fig_framework}
\end{figure*}

\subsection{Preliminaries}

\textbf{State-Space Model.} SSM is a linear time-invariant system that maps an input sequence $x(t)\in \mathbb{R}^N$ to an output sequence $y(t)\in \mathbb{R}^N$. They are mathematically represented by linear ordinary differential equations:
\begin{equation}
    h'(t)= \bm{\mathrm{A}}h(t)+ \bm{\mathrm{B}}x(t),
\end{equation}
\begin{equation}    y(t)=\bm{\mathrm{C}}h(t)+\bm{\mathrm{D}}x(t),
\end{equation}
where $h(t) \in \mathbb{R}^{N}$ indicates a hidden state, $h^{\prime}(t) \in \mathbb{R}^{N}$ refers to the time derivative of $h(t)$, and
$N$ represents the number of states.
Additionally, $\mathbf{A} \in \mathbb{R}^{N \times N}$ is the state transition matrix, $\mathbf{B} \in \mathbb{R}^{N\times 1}$ , $\mathbf{C} \in \mathbb{R}^{N \times 1}$ are projection matrices, and $\mathbf{D} \in\mathbb{R}^{N\times1}$ is served as a residual connected operation.

SSMs are continuous-time models, and it is challenging for them to be incorporated into deep learning networks. To solve the problem, discrete versions of SSMs are proposed, and the ordinary differential equations can be discretized by the zeroth-order hold rule. A timescale parameter $\Delta$ is incorporated to convert the continuous parameters $\bm{\mathrm{A}}$, $\bm{\mathrm{B}}$ into discrete parameters $\bm{\overline{\mathrm{A}}}$, $\bm{\overline{\mathrm{B}}}$ , respectively, as:
\begin{equation}
  \bm{\overline{\mathrm{A}}} = \mathrm{exp}({\Delta\bm{\mathrm{A}}}),
\end{equation}
{\begin{equation}
  \bm{\overline{\mathrm{B}}} = (\Delta \bm{\mathrm{A}})^{-1}(\mathrm{exp}({\Delta\bm{\mathrm{A}}})- \mathrm{\bm{I}}) \cdot \Delta \bm{\mathrm{B}}\approx\Delta \bm{\mathrm{B}}.
\end{equation}
}
{In practice, as noted in \cite{Mamba}, the projection matrix $\overline{\mathbf{B}}$ can be approximated by applying a first-order Taylor expansion to the term involving the matrix exponential.} After discretization, the ordinary differential equations of SSMs can be represented as follows: 
\begin{equation}
\begin{aligned}
  h_{t} & = \bm{\overline{\mathrm{A}}}h_{t-1} + \bm{\overline{\mathrm{B}}}x_{t} \; \\  \label{ssm}
  y_{t} & = \bm{\mathrm{C}}h_{t} + \bm{\mathrm{D}}x_{t} \; \\
\end{aligned}
\end{equation}

\textbf{Selective Scan Mechanism.} Traditional SSMs use a linear time-invariant framework, which means that the projection matrices remain fixed and unaffected by variations in the input sequence. However, this static configuration results in a lack of attention on individual elements within the sequence. To alleviate this limitation, Mamba \cite{Mamba} proposes a solution where the parameter matrices become input-dependent. In this way, SSMs can better manage complex sequences, potentially enhancing their capability through the transformation into linear time-varying systems.

\subsection{Overall Framework of the Proposed MSFMamba}

As depicted in Fig. \ref{fig_framework}, our MSFMamba consists of two parts: 1) The main network with a series of $L$ spatial-spectral Mamba modules to learn high-quality multi-source feature representations. 2) A classifier with two fully-connected layers for land cover classification. The number of spatial-spectral Mamba module $L$ is a critical parameter that will be discussed in the experiments.

The HSI is first handled by Principal Component Analysis (PCA) to select the best $N_p$ spectral bands. The HSI and LiDAR/SAR data are fed into the spatial-spectral Mamba modules for feature extraction and cross-modal feature fusion. The spatial-spectral Mamba modules are repeated $L$ times, and the final features are concatenated and fed into the classifier for classification. The classifier consists of two fully-connected layers. 

The Spatial-Spectral Mamba Module is the critical component in our MSFMamba. Details of the Spatial-Spectral Mamba Module are shown in Fig. \ref{fig_framework}{\color{blue}(a)}. The input HSI and LiDAR/SAR features are first fed into the {MSpa-Mamba} block for spatial feature extraction. To further exploit the spectral features of HSI, the Spe-Mamba block is employed for HSI feature extraction. Then, we obtain optimized features $\mathbf{F}_h$ and $\mathbf{F}_x$. Next, $\mathbf{F}_h$ and $\mathbf{F}_x$ are fed into the Fusion Mamba block for cross-modal feature fusion and refinement. Finally, the refined HSI feature $\mathbf{F}'_h$ and LiDAR/SAR feature $\mathbf{F}'_x$ are generated. 

As can be observed from Fig. \ref{fig_framework}{\color{blue}(a)}, the { MSpa-Mamba block}, Spe-Mamba block, and Fusion Mamba block are three key components in the Spatial-Spectral Mamba Module. We will give detailed descriptions of the three blocks in the following subsections.

\subsection{MSpa-Mamba Block}

\textbf{Overview of the MSpa-Mamba.}
The Mamba can model long-range feature dependencies, which is critical for understanding the global context in remote sensing images. Existing SSM-based models commonly use multi-scan strategy to ensure that every part of the image can establish connections with other parts. However, the multi-scan strategy significantly increases the feature redundancy of SSM. 

To solve the problem, we design a simple yet effective MSpa-Mamba block. As shown in Fig. \ref{fig_framework}{\color{blue}(b)}, the input feature $\mathbf{X}$ $\in$ $\mathbb{R}^{H\times W\times C}$ is fed into two parallel branches. In the first branch, the feature is processed by a linear layer, followed by a depth-wise convolution (DWConv), {Sigmoid Linear Unit (SiLU)} activation function, together with the { Multi-scale Spatial SSM (MSpa-SSM)} layer and LayerNorm. In the second branch, the feature is processed by a linear layer followed by the SiLU activation function. After that, features from the two branches are aggregated with element-wise multiplication. The computation of MSpa-Mamba block is as follows:
\begin{equation}
    \mathbf{X}_1=\textrm{LN}(\textrm{MSpa-SSM}(\textrm{SiLU}(\textrm{DWConv}(\textrm{Linear}(\mathbf{X}))))),
\end{equation}
\begin{equation}
    \mathbf{X}_2 = \textrm{SiLU}(\textrm{Linear}(\mathbf{(X)})),
\end{equation}
\begin{equation}
    \mathbf{X}_{out}=\textrm{Linear}(\mathbf{X}_1 \odot \mathbf{X}_2),
\end{equation}
where LN denotes the LayerNorm, and $\odot$ denotes the element-wise multiplication.

\textbf{MSpa-SSM.}
As shown in Fig. \ref{fig_framework}{\color{blue}(b)}, MSpa-SSM is the critical part of the MSpa-Mamba block. In MSpa-SSM, features at multiple scales are generated through DWConv with different stride values. These multi-scale feature maps are then handled by four scanning routes within SSM \cite{ly24vmamba}. The scanning routes are divided into two groups: two maintain the original resolution and are processed by the SSM, while the others are downsampled, processed by the SSM, and then upsampled. This strategy reduces the overall feature volume by generating more compact feature representations, effectively preventing the accumulation of redundant information across all scanning routes. To be more specific, we use DWConv with strides of 1 and 2 to obtain feature map $\mathbf{Z}_1\in \mathbb{R}^{H\times W\times C}$ and $\mathbf{Z}_2 \in \mathbb{R}^{\frac{H}{2}\times \frac{W}{2} \times C}$, respectively. Next, $\mathbf{Z}_1$ and $\mathbf{Z_2}$ are fed into the SSM as follows:
\begin{equation}
    [\mathbf{Y}_1, \mathbf{Y}_2]= \textrm{SSM}(\sigma_1(\mathbf{Z}_1), \sigma_2(\mathbf{Z}_1)),
\end{equation}
\begin{equation}
    [\mathbf{Y}_3, \mathbf{Y}_4]= \textrm{SSM}(\sigma_3(\mathbf{Z}_2), \sigma_4(\mathbf{Z}_2)),
\end{equation}
where $\sigma$ represents the transformation that reshapes the input features from $H \times W \times C$ to $HW \times C$, where $HW$ specifies the length of the spatial sequence to be processed by the SSM module. As illustrated in Fig. \ref{fig_framework}, the input is scanned in four distinct directions: the first scans rows first and then columns, while the second scans columns first and then rows. The third and fourth directions are the reverse of the first and second, respectively. Consequently, $\mathbf{Y}$ denotes the resulting processed sequence.

The obtained sequences are converted back into 2D feature maps, and the downsampled feature maps are interpolated for merging as follows:
\begin{equation}
    \mathbf{Z}'_i = \beta_i(\mathbf{Y}_i), ~~ i\in\{1,2,3,4\},
\end{equation}
\begin{equation}
    \mathbf{Z}'=\mathbf{Z}'_1 + \mathbf{Z}'_2 + \textrm{Inter}(\mathbf{Z}'_3 + \mathbf{Z}'_4),
\end{equation}
{where $\beta$ is the inverse transformation of $\sigma$, reshaping the input data from $\mathbb{R}^{HW \times C}$ back to $\mathbb{R}^{H \times W \times C}$.} Here, $\mathbf{Z}'$ represents the feature map enhanced by the SSM module, and \textquoteleft Inter \textquoteright refers to the interpolation operation.

\subsection{Spe-Mamba Block}

HSI covers a much larger spectral range with a higher spectral resolution. Effectively modeling the spectral relationship is important. In this paper, we design a {Spe-Mamba} block for spectral feature modeling. 

Details of the Spe-Mamba are illustrated in Fig. \ref{fig_framework}{\color{blue}(c)}. The input feature is fed into two parallel branches. In the first branch, the feature is handled by a linear layer, DWConv, and SiLU activation, together with the Spectral SSM (Spe-SSM) layer. In the second branch, the feature is handled by a linear layer and SiLU activation. Finally, features from both branches are aggregated with element-wise multiplication.

The Spe-Mamba shares a similar structure with the MSpa-Mamba, with the primary distinction being the SSM layer. In MSpa-SSM, features are transformed into 1D sequences along the spatial dimension, while in Spe-Mamba, they are converted into 1D sequences along the channel dimension, with scanning limited to two directions. Before being fed into Spe-SSM, the feature dimensions are reshaped from $\mathbb{R}^{H \times W \times C}$ to $\mathbb{R}^{C \times HW}$, where $C$ specifies the length of the sequence processed by the SSM module.

\begin{figure}[] 
\centering
\subfigure[Fusion Mamba (Fus-Mamba) block ]{
\includegraphics [width=2.5in]{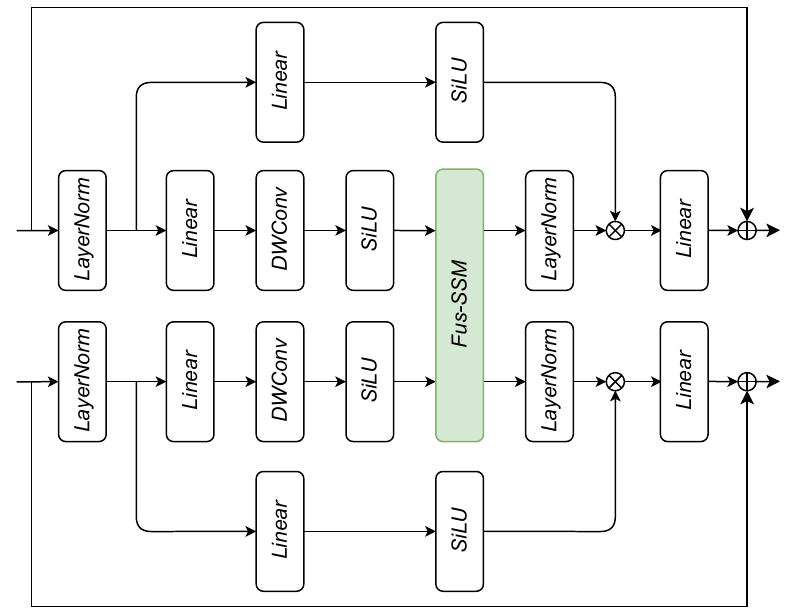}
\label{fig_fusa}}
\subfigure[Fusion SSM (Fus-SSM)]{
\includegraphics [width=3in]{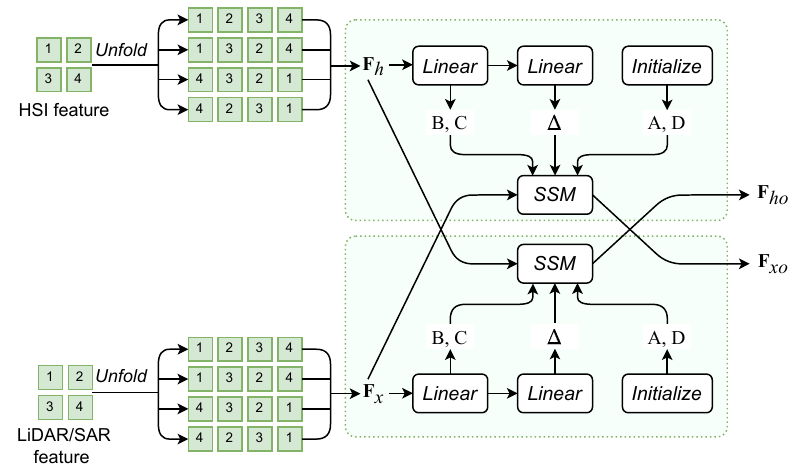}
\label{fig_fusb}}
\caption{Illustration of Fusion Mamba (Fus-Mamba) block, as well as the Fusion SSM (Fus-SSM).}
\label{fig_fus}
\end{figure}

\subsection{Fus-Mamba Block}

Traditional SSMs typically handle single inputs and struggle to effectively integrate cross-modal information from multi-source data. To address this limitation, we propose the Fus-Mamba block, which integrates features from different modalities. As shown in Fig. \ref{fig_fus}, HSI and LiDAR/SAR features undergo a similar processing flow to that of MSpa-Mamba and Spe-Mamba. Multi-source features are then fed into the Fusion-SSM (Fus-SSM) block for cross-modal feature fusion and interaction.

The Fus-SSM serves as the core of Fus-Mamba, with its structure detailed in Fig. \ref{fig_fusb}. Fus-SSM is designed symmetrically. In the top branch, the sequenced HSI feature $\mathbf{F}_h \in \mathbb{R}^{P \times C}$ generates projection and timescale parameters, while the sequenced LiDAR/SAR feature $\mathbf{F}_x \in \mathbb{R}^{P \times C}$ is processed to produce the output feature $\mathbf{F}_{xo}$. The bottom branch mirrors this structure: here, $\mathbf{F}_x$ generates the projection and timescale parameters, and $\mathbf{F}_h$ is processed to generate the output feature $\mathbf{F}_{ho}$.

Algorithm \ref{alg_fus} outlines the computation process for $\mathbf{F}_{ho}$. Parameters A and D are initialized at the start, while B, C, and $\Delta$ are derived from linear layers applied to the sequence $\textbf{F}_x$ from another modality. In the original Mamba model, B, C, and $\Delta$ are derived from the sequence being processed, whereas in Fus-SSM, B, C, and $\Delta$ come from a different modality, specifically $\textbf{F}_x$. Since the computation for $\mathbf{F}_{xo}$ closely resembles that of $\mathbf{F}_{ho}$, it is omitted for brevity.

\textbf{Comparison with Cross Mamba Block (CroMB)}:
Although the Fus-Mamba Block shares conceptual similarities with the Cross Mamba Block (CroMB) in Sigma\cite{Sigma}, it incorporates a more integrated cross-modal design. In CroMB, components $\mathbf{\overline{A}}$ and $\mathbf{\overline{B}}$ are generated solely from the sequence being processed, with only $\mathbf{C_X}$ relying on the sequence from the alternate modality.
In contrast, our Fus-SSM utilizes features from the alternative modality to generate all three parameters ($\mathbf{\overline{A}}$, $\mathbf{\overline{B}}$ and $\mathbf{C}$) when processing a given sequence. This approach enables the Fus-Mamba Block to capture deeper interactions and dependencies between modalities, fully leveraging cross-modal information to improve feature alignment and enhance overall model performance.

\begin{algorithm}[t]
\caption{The computation of $\mathbf{F}_{ho}$ in Fig. \ref{fig_fusb}}
\label{alg_fus}
\renewcommand{\algorithmicrequire}{\textbf{Input:}}
\renewcommand{\algorithmicensure}{\textbf{Output:}}
\begin{algorithmic}[1]
\Require $\mathbf{F}_h, \mathbf{F}_x :\rm{\textcolor[RGB]{0, 100, 0}{(P, C), (P, C)}}$
\Ensure $\mathbf{F}_{ho}:\rm{\textcolor[RGB]{0, 100, 0}{(P, C)}}$
\State  $\mathbf{A}:{\rm{\textcolor[RGB]{0, 100, 0}{(C, N)}}}\gets \mathbf{Parameter}$
\State  $\mathbf{D}:{\rm{\textcolor[RGB]{0, 100, 0}{(C)}}}\gets \mathbf{Parameter}$ 
\Statex \textcolor{gray}{ /* A and D are randomly initialized */}
\State  $\mathbf{B}:{\rm{\textcolor[RGB]{0, 100, 0}{(P, N)}}}\gets \mathbf{Linear}(\mathbf{F}_x)$ 
\State  $\mathbf{C}:{\rm{\textcolor[RGB]{0, 100, 0}{(P, N)}}}\gets \mathbf{Linear}(\mathbf{F}_x)$ 
\State  \smash{$\mathbf{\Delta}:{\rm{\textcolor[RGB]{0, 100, 0}{(P, C)}}}\gets \log(1+\exp(\mathbf{Linear}(\mathbf{F}_x)+\mathbf{Parameter})$} 
\Statex \textcolor{gray}{/* The selection mechanism */}
\State $\mathbf{\overline{A}}:{\rm{\textcolor[RGB]{0, 100, 0}{(P, C, N)}}}\gets \rm{exp(\mathbf{\Delta}\otimes \mathbf{A})}$ 
\State $\mathbf{\overline{B}}:\rm{\textcolor[RGB]{0, 100, 0}{(P, C, N)}}\gets \mathbf{\Delta}\otimes \mathbf{B}$ 
\State $\mathbf{F}_{ho} \gets {\rm{SSM}(\mathbf{\overline{A}}, \mathbf{\overline{B}}, \mathbf{C},\mathbf{D})}(\mathbf{F}_h)$ 
\Statex \textcolor{gray}{/* SSM denotes Eq. \ref{ssm} implemented by selective scan */}
\Statex  $\mathbf{return}\;\mathbf{F}_{ho}$
\end{algorithmic}
\end{algorithm}

\section{Experimental Results and Analysis}

\definecolor{m1}{HTML}{1AA319}
\definecolor{m2}{HTML}{D8D8D8}  
\definecolor{m3}{HTML}{D85959}  
\definecolor{m4}{HTML}{00CC33}             
\definecolor{m5}{HTML}{CC9934}  
\definecolor{m6}{HTML}{F4E701}  
\definecolor{m7}{HTML}{CC66CC}  
\definecolor{m8}{HTML}{0035FF} 

\definecolor{mm1}{HTML}{1AA319}
\definecolor{mm2}{HTML}{D8D8D8}
\definecolor{mm3}{HTML}{D85959}
\definecolor{mm4}{HTML}{00CC33}
\definecolor{mm5}{HTML}{F4E701}
\definecolor{mm6}{HTML}{CC66CC}
\definecolor{mm7}{HTML}{0035FF}

\definecolor{hh1}{HTML}{32CD33}  
\definecolor{hh2}{HTML}{ADFF30} 
\definecolor{hh3}{HTML}{008081}  
\definecolor{hh4}{HTML}{228B22}  
\definecolor{hh5}{HTML}{2E4F4E}  
\definecolor{hh6}{HTML}{8B4512}  
\definecolor{hh7}{HTML}{00FFFF}  
\definecolor{hh8}{HTML}{FFFFFF}  
\definecolor{hh9}{HTML}{D3D3D3}  
\definecolor{hh10}{HTML}{FE0000}
\definecolor{hh11}{HTML}{A9A9A9}
\definecolor{hh12}{HTML}{696969}
\definecolor{hh13}{HTML}{8B0001}
\definecolor{hh14}{HTML}{C86400}
\definecolor{hh15}{HTML}{FEA500}
\definecolor{hh16}{HTML}{FFFF00}
\definecolor{hh17}{HTML}{DAA521}
\definecolor{hh18}{HTML}{FF00FE}
\definecolor{hh19}{HTML}{0000FE}
\definecolor{hh20}{HTML}{3FE0D0}

\definecolor{hn1}{HTML}{000083}
\definecolor{hn2}{HTML}{0000CB}
\definecolor{hn3}{HTML}{0013FF}
\definecolor{hn4}{HTML}{005BFF}
\definecolor{hn5}{HTML}{00A7FF}
\definecolor{hn6}{HTML}{00EFFF}
\definecolor{hn7}{HTML}{37FFC7}
\definecolor{hn8}{HTML}{83FF7B}
\definecolor{hn9}{HTML}{CBFF33}
\definecolor{hn10}{HTML}{FFEB00}
\definecolor{hn11}{HTML}{FFA300}
\definecolor{hn12}{HTML}{FF5700}
\definecolor{hn13}{HTML}{FF0F00}
\definecolor{hn14}{HTML}{C70000}
\definecolor{hn15}{HTML}{7F0000}

\subsection{Dataset Description}

To assess the effectiveness of our proposed MSFMamba, we applied it on four multi-source remote sensing datasets: Berlin, Augsburg, Houston 2018, and Houston 2013. The Berlin and Augsburg datasets are used for hyperspectral and SAR data classification, while the Houston 2013 and Houston 2018 datasets are used for hyperspectral and LiDAR data classification.

\begin{table}[htbp]
\renewcommand\arraystretch{1}
\centering
\caption{The number of training and test samples for the Berlin, Augsburg, Houston2018 and Houston2013 datasets.}
\scalebox{0.9}{
\begin{tabular}{cccc|cc}
\multicolumn{6}{c}{Berlin dataset} \\
\hline\toprule
    No. & Name & Color & ~ & Train & Test \\
\midrule
    1 & Forest & \cellcolor{m1} & & 443 & 54511\\
    2 & Residential area & \cellcolor{m2} & & 423 & 268219 \\
    3 & Industrial area & \cellcolor{m3} & & 499 & 19067\\
    4 & Low plants & \cellcolor{m4} & & 376 & 58906 \\
    5 & Soil & \cellcolor{m5} & & 331 & 17095 \\
    6 & Allotment & \cellcolor{m6} & & 280 & 13025\\
    7 & Commercial area & \cellcolor{m7} & & 298 & 24526\\
    8 & Water & \cellcolor{m8} & & 170 & 6502\\
\midrule
    \multicolumn{3}{c}{Total} & & 2820 & 461851\\
\bottomrule\hline
\multicolumn{6}{c}{~~~~}\\

\multicolumn{6}{c}{Augsburg dataset} \\
\hline\toprule
    No. & Name & Color & ~ & Train & Test \\
\midrule
    1 & Forest & \cellcolor{mm1} & & 146 & 13361\\
    2 & Residential area & \cellcolor{mm2} & & 264 & 30065 \\
    3 & Industrial area & \cellcolor{mm3} & & 21 & 3830\\
    4 & Low plants & \cellcolor{mm4} & & 248 & 26609 \\
    5 & Allotment & \cellcolor{mm5} & & 52 & 523\\
    6 & Commercial area & \cellcolor{mm6} & & 7 & 1638\\
    7 & Water & \cellcolor{mm7} & & 23 & 1507\\
\midrule
    \multicolumn{3}{c}{Total} & & 761 & 77533\\
\bottomrule\hline
\multicolumn{6}{c}{~~~~}\\

\multicolumn{6}{c}{Houston2018 dataset} \\
\hline\toprule
    No. & Name & Color & ~ & Train & Test \\
\midrule
    1 & Health grass & \cellcolor{hh1} & & 1000 & 38196\\
    2 & Stressed grass & \cellcolor{hh2} & & 1000 & 129008\\
    3 & Artificial turf & \cellcolor{hh3} & & 1000 & 1736\\
    4 & Evergreen trees & \cellcolor{hh4} & & 1000 & 53322\\
    5 & Deciduous trees & \cellcolor{hh5} & & 1000 & 19172\\
    6 & Bare earth & \cellcolor{hh6} & & 1000 & 17064\\
    7 & Water & \cellcolor{hh7} & & 500 & 564\\
    8 & Residential buildings & \cellcolor{hh8} & & 1000 & 157995\\
    9 & Non-residential buildings & \cellcolor{hh9} & & 1000 & 893769\\
    10 & Roads & \cellcolor{hh10} & & 1000 & 182283\\
    11 & Sidewalks & \cellcolor{hh11} & & 1000 & 135035\\
    12 & Crosswalks & \cellcolor{hh12} & & 1000 & 5059\\
    13 & Major thoroughfares & \cellcolor{hh13} & & 1000 & 184438\\
    14 & Highways & \cellcolor{hh14} & & 1000 & 38438\\
    15 & Railways & \cellcolor{hh15} & & 1000 & 26748\\
    16 & Paved parking lots & \cellcolor{hh16} & & 1000 & 44932\\
    17 & Unpaved parking lots & \cellcolor{hh17} & & 250 & 337\\
    18 & Cars & \cellcolor{hh18} & & 1000 & 25289\\
    19 & Trains & \cellcolor{hh19} & & 1000 & 20479\\
    20 & Stadium seats & \cellcolor{hh20} & & 1000 & 26296\\
\midrule
    \multicolumn{4}{c|}{Total} & 18750 & 2000160\\
\bottomrule\hline
\multicolumn{6}{c}{~~~~}\\

\multicolumn{6}{c}{Houston2013 dataset} \\
\hline\toprule
    No. & Name & Color & ~ & Train & Test \\
\midrule
    1 & Health grass & \cellcolor{hn1} & & 198 & 1053\\
    2 & Stressed grass & \cellcolor{hn2} & & 190 & 1064\\
    3 & Synthetic grass & \cellcolor{hn3} & & 192 & 505\\
    4 & Trees & \cellcolor{hn4} & & 188 & 1056\\
    5 & Soil & \cellcolor{hn5} & & 186 & 1056\\
    6 & Water & \cellcolor{hn6} & & 182 & 143\\
    7 & Residential & \cellcolor{hn7} & & 196 & 1072\\
    8 & Commercial & \cellcolor{hn8} & & 191 & 1053\\
    9 & Road & \cellcolor{hn9} & & 193 & 1059\\
    10 & Highway & \cellcolor{hn10} & & 191 & 1036\\
    11 & Railway & \cellcolor{hn11} & & 181 & 1054\\
    12 & Parking lot 1 & \cellcolor{hn12} & & 192 & 1041\\
    13 & Parking lot 2 & \cellcolor{hn13} & & 184 & 285\\
    14 & Tennis court & \cellcolor{hn14} & & 181 & 247\\
    15 & Running track & \cellcolor{hn15} & & 187 & 473\\
\midrule
    \multicolumn{4}{c|}{Total} & 2832 & 12197\\
\bottomrule\hline
\label{table_dataset}
\end{tabular}}
\end{table}

\begin{figure*}[ht]
\centering
\includegraphics [width=0.99\textwidth]{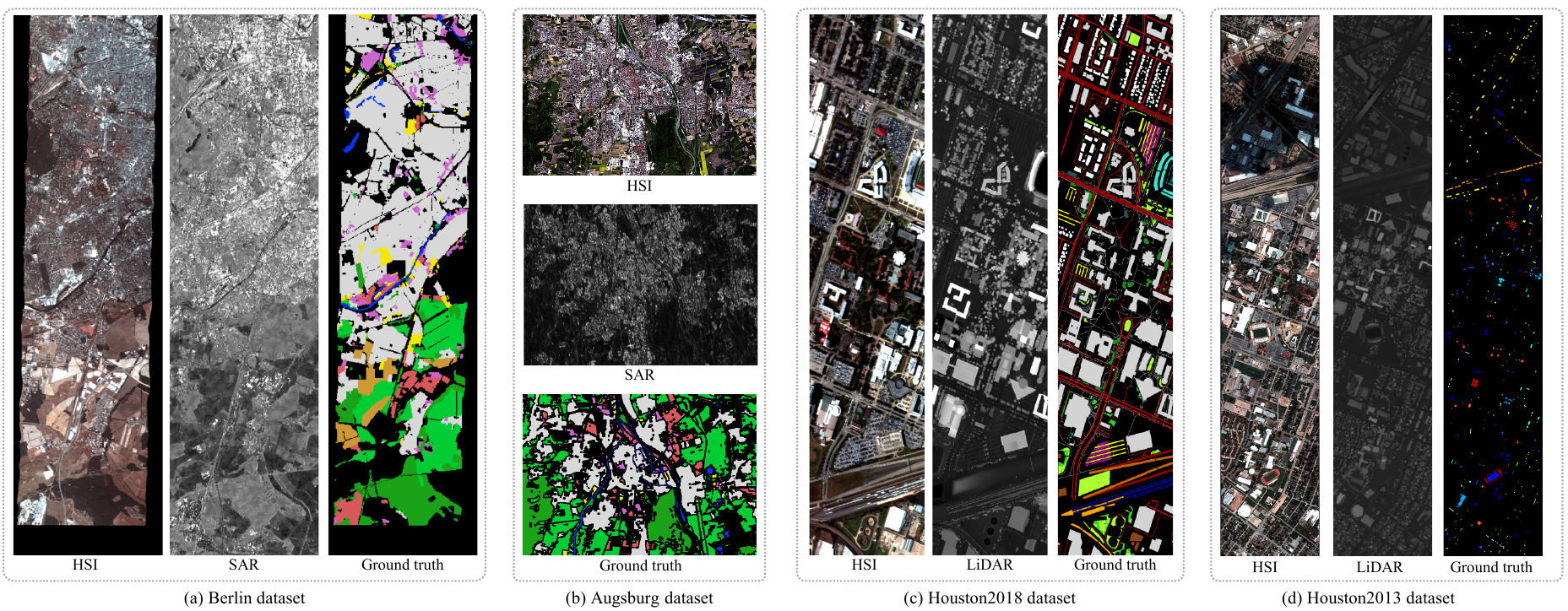}
\caption{Visualization of the four datasets used in our experiments.}
\label{dataset}
\end{figure*}

\textbf{Berlin Dataset.} It covers the urban and rural areas of Berlin. The hyperspectral data is simulated { Environmental Mapping and Analysis Program(EnMAP) data based on Hyperspectral Mapper(HyMap) data}. The SAR data is Sentinel-1 dual-Pol (VV–VH) single look complex (SLC) product obtained from the European Space Agency (ESA) \cite{ESA}. The HSI has 797$\times$220 pixels, 244 spectral bands in the wavelength range of 400–2500 nm \cite{berlin}. The SAR image has 1723$\times$476 pixels. The nearest neighbor interpolation was employed to match the spatial resolution of the HSI and SAR data.

\textbf{Augsburg Dataset.} This dataset is composed of HSI and SAR data from Augsburg, Germany. The HSI was acquired by the HySpex sensor, and the SAR data was acquired by the Sentinel-1 sensor. All images are with the ground sample distance (GSD) of 30 m. The spatial size of both images is 332$\times$485 pixels. The HSI contains 180 spectral bands ranging from 0.4 to 2.5 $\mu$m, and the SAR data has four features derived from polarization decomposition (VV intensity, VH intensity, real part and imaginary part of the off-diagonal element of the PolSAR covariance matrix).

\textbf{Houston2018 Dataset.} This dataset covers the University of Houston campus and the neighboring urban area. The HSI contains 48 bands in the wavelength range of 380–1050 nm. LiDAR data is a multispectral LiDAR data with three bands. The University of Houston released the dataset as part of the 2018 IEEE GRSS Data Fusion Contest \cite{ls18grsm}. We use the training subset of the whole dataset in this paper.

\textbf{Houston2013 Dataset.} The Houston2013 dataset, part of the 2013 IEEE Geoscience and Remote Sensing Society Data Fusion Contest, provides a unique perspective on urban land cover in Houston, Texas, and its surrounding areas. Captured at a spatial resolution of 2.5 meters, the dataset includes hyperspectral imagery with 144 spectral bands covering wavelengths from 380 to 1050 nm, along with LiDAR data that provides precise elevation information. Together, these data sources encompass 15 distinct land cover types, offering valuable insights for remote sensing and urban analysis.

Table \ref{table_dataset} enumerates the number of samples for both training and testing on the four datasets. Fig. \ref{dataset} displays the HSI data in false color, along with the LiDAR/SAR images and the ground truth. The evaluation metrics used in this paper include Overall Accuracy(OA), Average Accuracy(AA), and Kappa. OA represents the proportion of correctly classified samples out of all the samples in the dataset, providing a measure of the model's overall performance. 
AA is the average recall rate across all classes, reflecting the model's effectiveness in handling different categories, particularly in imbalanced datasets. Kappa, on the other hand, is a statistical measure that quantifies the agreement between the predicted and true labels, adjusted for random chance, offering a more robust evaluation of classifier performance than simple accuracy, especially when dealing with imbalanced or noisy data.


\begin{figure}[htb]
    \centering
    \includegraphics[width=2in]{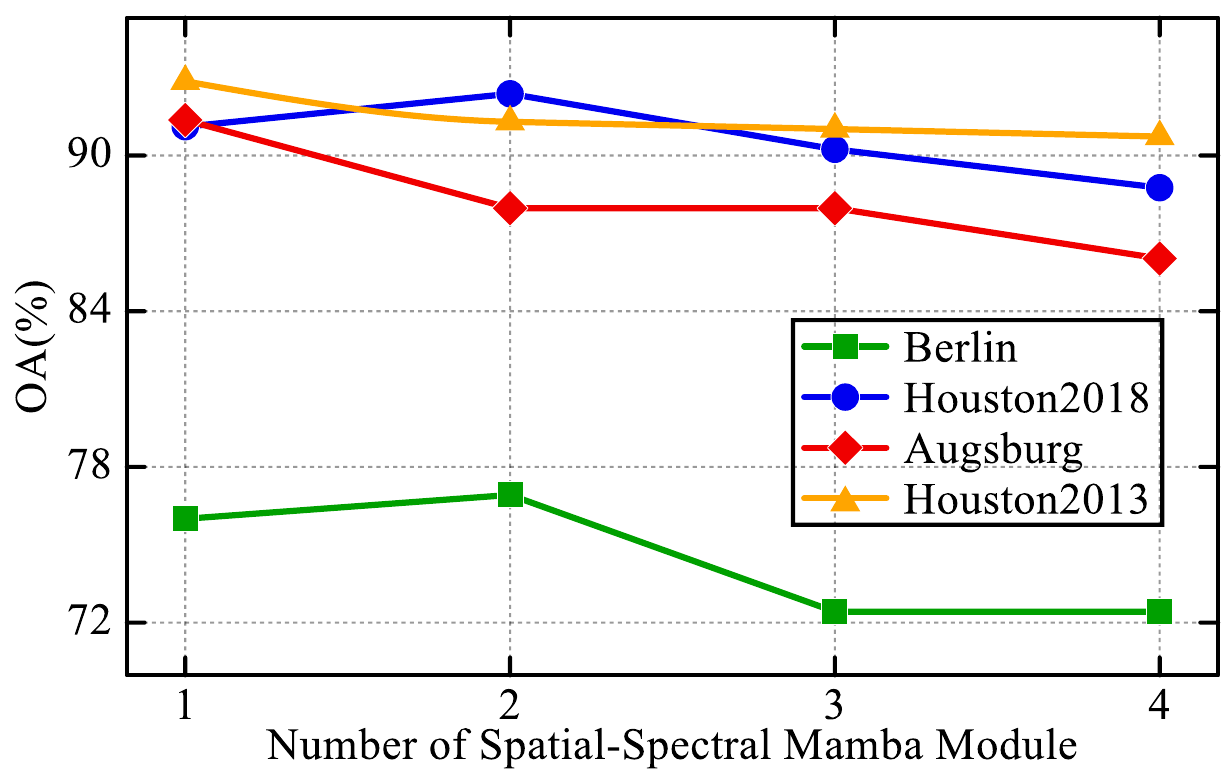}
    \caption{{The relationship between OA and the number of spatial-spectral Mamba module.}}
    \label{fig_para_ssmm}
\end{figure}

\begin{figure}[htb]
    \centering
    \includegraphics[width=2in]{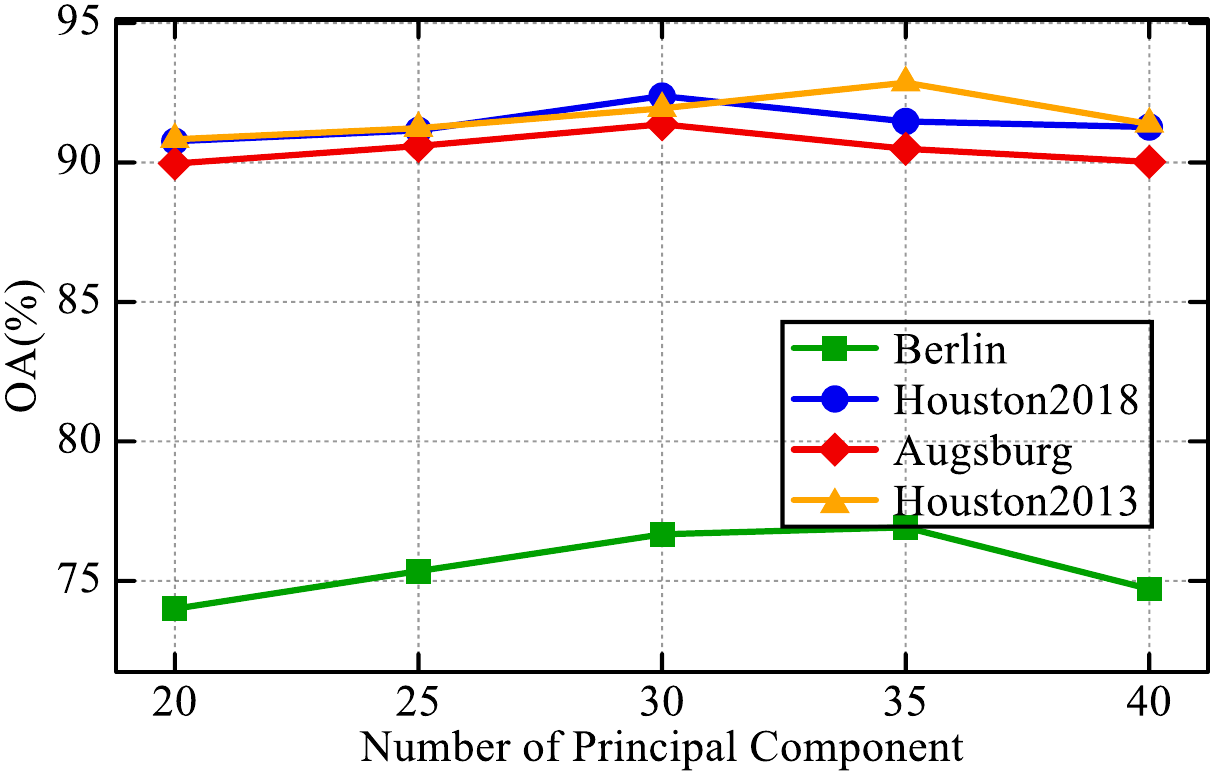}
    \caption{The relationship between OA and the number of principal components in HSI.}
    \label{fig_para_pca}
\end{figure}

\begin{figure}[htb]
    \centering
    \includegraphics[width=2in]{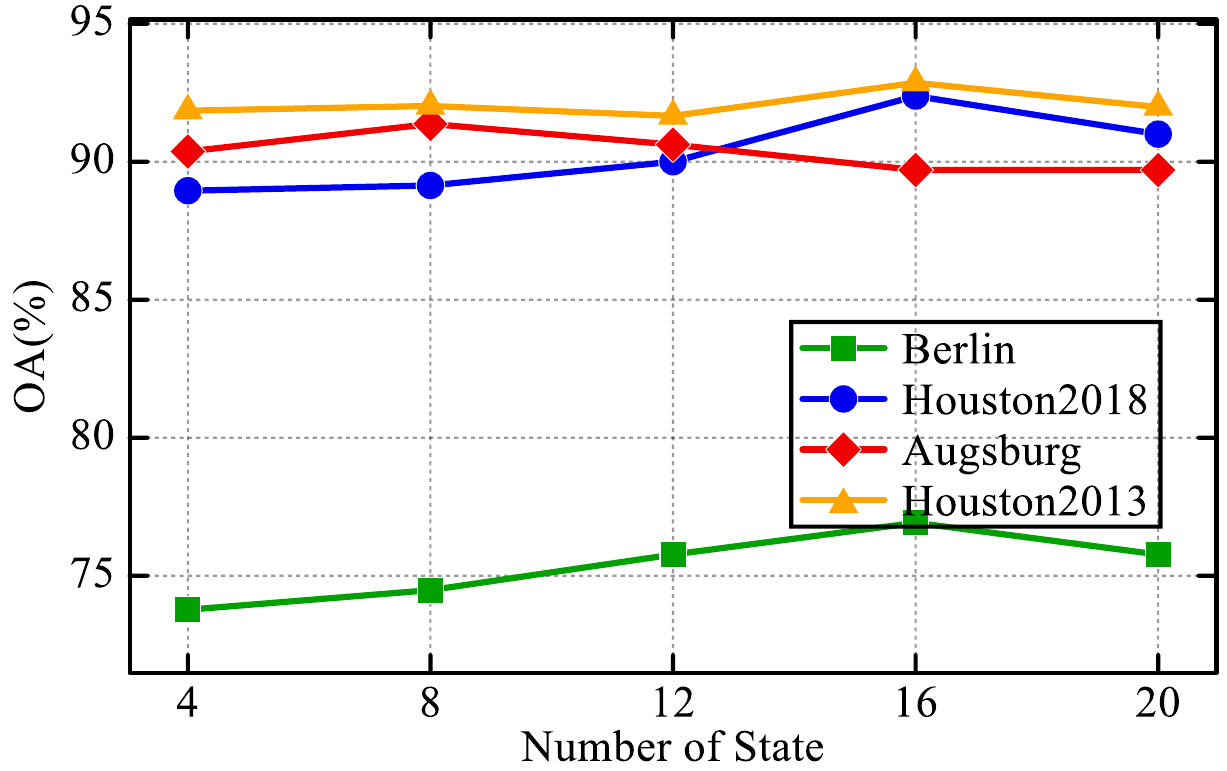}
    \caption{The relationship between OA and the number of states in SSM.}
    \label{fig_para_state}
\end{figure}

\subsection{Parameter Analysis}

We provide a detailed analysis of several important parameters that may affect the classification performance of our MSFMamba. These parameters include the number of spatial-spectral Mamba module, the number of principal components in HSI, and the number of state in SSM. 

\textbf{Number of Spatial-Spectral Mamba Module.} The spatial-spectral Mamba module is the critical part of our MSFMamba, and the number of modules, $L$, is a key parameter that affects classification performance. We tested the optimal number of spatial-spectral Mamba modules for each dataset. Fig. \ref{fig_para_ssmm} illustrates the relationship between the OA value and $L$ on each dataset. Our MSFMamba achieves the best classification performance on the Augsburg and Houston2013 datasets when $L=1$. For the Berlin and Houston2018 datasets, the best performance is observed when $L=2$. Therefore, in our subsequent experiments, we set $L=1$for the Augsburg and Houston2013 datasets and $L=2$ for the Berlin and Houston2018 datasets.

\textbf{Number of Principal Components in HSI.} We use PCA to reduce the spectral redundancy of the input HSI, with the number of principal components,$N_p$, being a critical parameter. We tested different values of $N_p$ ranging from 20 to 40, and the experimental results are shown in Fig. \ref{fig_para_pca}. It can be seen that when $N_p=30$, the OA values for the Augsburg and Houston2018 datasets reach their highest. For the Berlin and Houston2013 datasets, the best OA values are achieved when $N_p=35$.

\textbf{Number of States in SSM.} The number of states is an important parameter in the state space model. We tested five different values for the number of states, ranging from 4 to 20. As shown in Fig. \ref{fig_para_state}, the optimal number of states for the Houston2018, Houston2013, and Berlin datasets is 16, while for the Augsburg dataset, the optimal number of states is 8.

The best settings of these experiments were used to evaluate our MSFMamba for comparison with other state-of-the-art methods. All experiments were performed with an RTX 4090 GPU and 32GB of RAM using PyTorch.

\definecolor{n1}{RGB}{255,0,0}
\definecolor{n2}{RGB}{0,0,255}
\definecolor{n3}{RGB}{0,255,0}

\begin{table*}[htb]
\renewcommand\arraystretch{1.1}
\centering
\caption{Classification performance of different methods on the Berlin dataset.}
\scalebox{0.9}{
\begin{tabular}{c|ccccccccc}
\toprule
    ~~~Class~~~      & ~TBCNN~ & FusAtNet           & ~$S^2$ENet~ & ~DFINet~             & AsyFFNet         & ExViT       & HCT & MACN & MSFMamba\\\hline
    Forest             & ~81.75~ & ~86.24~           & ~83.27~ & ~82.04~             & ~\textbf{88.35}~         & ~84.71~    & ~83.18~ & ~82.21~ & ~82.39~ \\ 
    Residential area   & ~76.26~ & ~\textbf{91.38}~           & ~72.07~ & ~77.78~             & ~74.44~         & ~76.19~    & ~78.80~ & ~78.16~ & ~82.05~ \\ 
    Industrial area    & ~39.67~ & ~19.76~           & ~46.66~ & ~47.96~             & ~48.21~         & ~44.15~    & ~43.59~ & ~47.79~ & ~\textbf{48.91}~ \\
    Low plants         & ~49.78~ & ~20.00~           & ~72.08~ & ~77.78~             & ~71.41~         & ~77.96~    & ~76.48~ & ~78.04~ & ~\textbf{79.72}~ \\
    Soil               & ~\textbf{89.42}~ & ~48.72~           & ~77.94~ & ~87.85~             & ~80.75~         & ~67.13~    & ~82.92~ & ~77.26~ & ~85.22~ \\
    Allotment          & ~54.36~ & ~38.89~           & ~70.62~ & ~44.75~             & ~48.35~         & ~61.95~    & ~56.46~ & ~66.07~ & ~\textbf{71.25}~ \\
    Commercial area    & ~4.65~  & ~18.47~           & ~\textbf{36.48}~ & ~29.85~             & ~27.97~         & ~29.28~    & ~18.96~ & ~28.90~ & ~31.08~ \\
    Water              & ~41.93~ & ~29.61~           & ~54.64~ & ~60.09~             & ~\textbf{75.16}~         & ~56.98~    & ~44.16~ & ~50.22~ & ~38.48~ \\\hline
    OA                 & ~67.60~ & ~70.91~           & ~70.38~ & ~73.69~             & ~71.65~         & ~72.59~    & ~73.42~ & ~73.98~ & ~\textbf{76.92}~ \\ 
    AA                 & ~54.72~ & ~44.13~           & ~64.22~ & ~63.51~             & ~64.33~         & ~62.39~    & ~60.57~ & ~63.58~ & ~\textbf{64.88}~ \\ 
    Kappa              & ~50.96~ & ~51.07~           & ~57.73~ & ~61.02~             & ~58.74~         & ~59.72~    & ~60.29~ & ~61.19~ & ~\textbf{64.88}~ \\ 
\bottomrule
\end{tabular}}
\label{berlin_compare_table}
\end{table*}

\begin{figure*}[h!]
\centering
\includegraphics [width=0.95\textwidth]{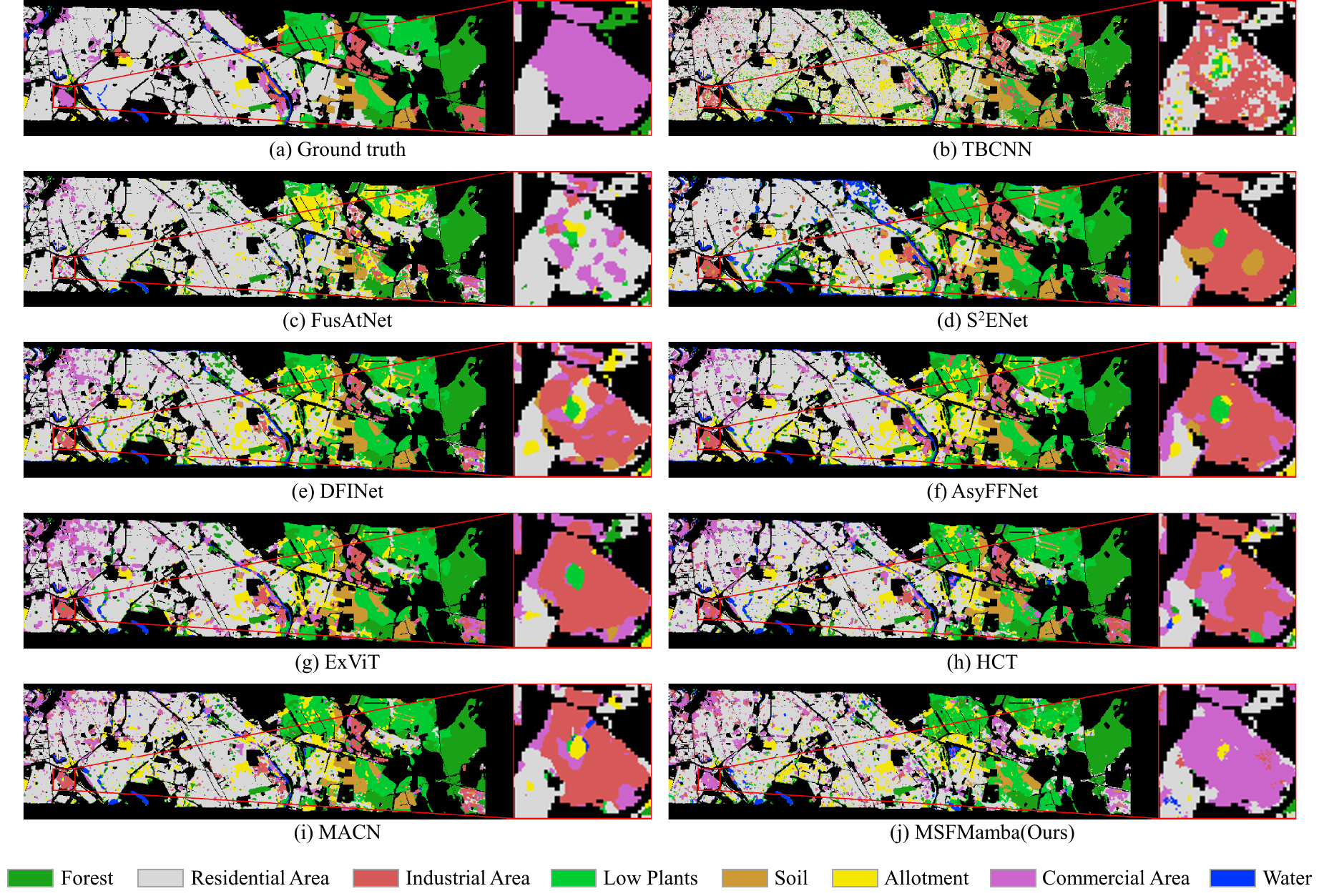}
\caption{Classification results of different methods for the Berlin dataset.}
\label{berlin_fig_com}
\end{figure*}

\definecolor{n1}{RGB}{255,0,0}
\definecolor{n2}{RGB}{0,0,255}
\definecolor{n3}{RGB}{0,255,0}

\begin{table*}[t]
\renewcommand\arraystretch{1.1}
\centering
\caption{Classification performance of different methods on the Augsburg dataset.}
\scalebox{0.9}{
\begin{tabular}{c|cccccccccc}
\toprule
	~~~Class~~~    & ~TBCNN~ & FusAtNet & ~$S^2$ENet~ & ~DFINet~        & AsyFFNet & ExViT & HCT    & MACN           & MSFMamba\\\hline
	Forest             & ~90.88~ & ~94.34~ & ~\textbf{98.10}~ & ~97.38~          & ~97.56~ & ~90.04~ & ~97.26~ & ~97.84~      & ~97.17~ \\ 
	Residential area   & ~93.89~ & ~92.56~ & ~99.08~ & ~98.37~ & ~\textbf{99.16}~ & ~95.44~ & ~98.63~ & ~98.73~ & ~98.15~ \\ 
    Industrial area    & ~8.28~  & ~47.70~ & ~12.19~ & ~61.31~     & ~\textbf{61.38}~ & ~34.58~ & ~36.61~ & ~43.94~   & ~50.26~ \\
    Low plants         & ~91.97~ & ~85.96~ & ~91.78~ & ~92.63~              & ~83.47~ & ~90.68~ & ~93.54~ & ~94.11~            & ~\textbf{95.52}~ \\
    Allotment          & ~38.24~ & ~49.33~ & ~45.12~ & ~49.33~     & ~42.64~ & ~51.82~ & ~51.43~ & ~52.39~     & ~\textbf{53.35}~ \\
    Commercial area    & ~1.40~  & ~11.52~ & ~1.22~  & ~3.54~              & ~5.58~  & ~\textbf{28.63}~ & ~7.97~ & ~6.80~   & ~2.63~ \\
    Water              & ~10.82~ & ~45.27~ & ~24.09~ & ~26.61~              & ~45.81~ & ~17.65~ & ~46.34~ & ~48.54~              & ~\textbf{49.97}~ \\\hline
    OA                 & ~84.53~ & ~85.31~ & ~88.22~ & ~90.66~              & ~88.25~ & ~86.65~ & ~90.33~ & ~91.06~              & ~\textbf{91.38}~ \\ 
    AA                 & ~47.93~ & ~62.93~ & ~53.08~ & ~61.31~              & ~62.23~ & ~58.41~ & ~61.68~ & ~63.19~              & ~\textbf{63.31}~ \\ 
    Kappa              & ~77.13~ & ~86.33~ & ~86.47~ & ~86.47~              & ~83.13~ & ~80.79~ & ~86.03~ & ~87.06~              & ~\textbf{87.45}~ \\ 
\bottomrule
\end{tabular}}
\label{augsburg_compare_table}
\end{table*}

\begin{figure*}[h!]
\centering
\includegraphics [width=0.95\textwidth]{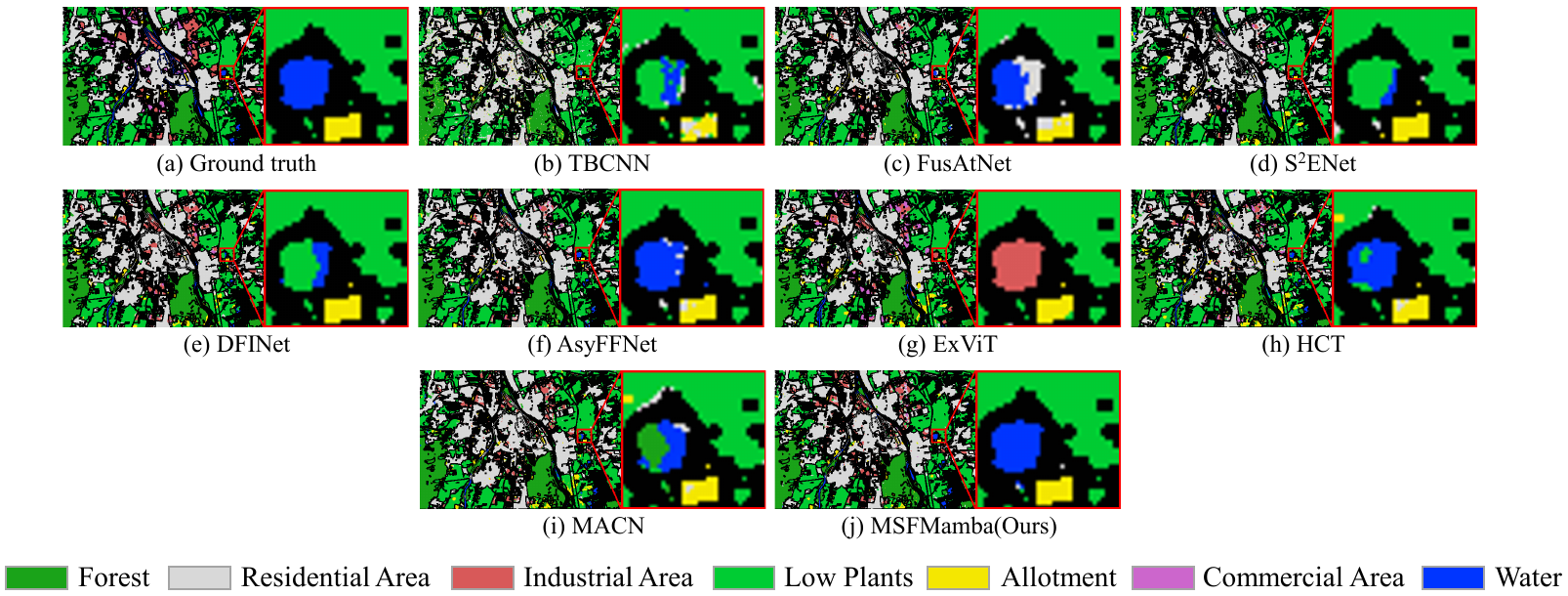}
\caption{Classification results of different methods for the Augsburg dataset.}
\label{augsburg_fig_com}
\end{figure*}

\begin{table*}[h!]
\renewcommand\arraystretch{1.1}
\centering
\caption{Classification performance of different methods on the Houston 2018 dataset.}
\scalebox{0.9}{
\begin{tabular}{c|cccccccccc}
\toprule
	Class                     & TBCNN & FusAtNet & $S^2$ENet & DFINet & AsyFFNet & ExViT & HCT & MACN & MSFMamba \\ \hline
	Health grass                & 94.84 & 60.69 & \textbf{99.15} & 94.29 & 96.9 & 93.34 & 97.38 & 89.53 & 90.38 \\ 
	Stressed grass              & 92.6  & 93.85 & 84.35 & 92.71 & 91.82 & 93.44 & 89.4  & 93.5  & \textbf{94.04} \\ 
	Artificial turf             & \textbf{100.0}   & \textbf{100.0}   & \textbf{100.0}   & \textbf{100.0}   & \textbf{100.0}  & \textbf{100.0}   & \textbf{100.0}   & \textbf{100.0}   & \textbf{100.0}   \\ 
	Evergreen trees             & 98.8  & 99.02 & 98.13 & 98.92 & 99.29 & 95.87 & 98.9  & 99.1  & \textbf{99.34} \\ 
	Deciduous trees             & 97.12 & 94.73 & 86.39 & 97.77 & \textbf{99.52} & 96.85 & 98.89 & 98.49 & 99.11 \\ 
	Bare earth                  & 99.61 & 99.87   & 99.33 & \textbf{100.0}   & 99.98 & \textbf{100.0}   & \textbf{100.0}   & 99.96 & \textbf{100.0}   \\ 
	Water                       & \textbf{100.0}   & \textbf{100.0}   & \textbf{100.0}   & \textbf{100.0}   & \textbf{100.0}  & \textbf{100.0}   & \textbf{100.0}   & \textbf{100.0}   & \textbf{100.0}   \\ 
	Residential buildings       & 93.21 & 95.51  & 97.17 & 97.36 & 95.16 & \textbf{97.39} & 95.3  & 95.31 & 96.27 \\ 
	Non-residential buildings   & 91.3  & 93.64 & 93.93 & 93.29 & 95.02 & 93    & 95.25 & 93.86 & \textbf{95.58} \\ 
	Roads                       & 61.06 & 70.94 & 70.17 & 75.73 & 72.72 & 72.18 & 72.77 & 72.11 & \textbf{75.9}  \\ 
	Sidewalks                   & 75.91 & 79.11 & 72.11 & \textbf{83.67} & 77.18 & 69.69 & 73.27 & 80.69 & 81.43 \\ 
	Crosswalks                  & 85.31 & 86.06 & 82.47 & 94.23 & 93.14 & 88.12 & \textbf{98.95} & 97.69 & 96.22 \\ 
	Major thoroughfares         & 72.77 & 76.12  & \textbf{89.44} & 81.2  & 84.67 & 86.12 & 80.9  & 79.05 & 86.89 \\ 
	Highways                    & 95.86 & 98.89 & 98.53 & 98.91 & \textbf{99.53} & 99.5  & 98.01 & 96.86 & 98.63 \\ 
	Railways                    & 99.78 & 99.88 & 99.21 & \textbf{99.94} & 99.90  & 99.87 & 99.74 & 99.9  & 99.69 \\ 
	Paved parking lots          & 90.74 & 95.88 & 95.62 & 98.37 & \textbf{99.18} & 97.02 & 97.07 & 97.74 & 98.94 \\ 
	Unpaved parking lots        & \textbf{100.0}   & \textbf{100.0}   & \textbf{100.0}   & \textbf{100.0}   & \textbf{100.0}  & \textbf{100.0}   & \textbf{100.0}   & \textbf{100.0}   & \textbf{100.0}   \\ 
	Cars                        & 98.47 & 94.25 & 96.77 & \textbf{99.09} & 96.64 & 98.13 & 99.01 & 98.74 & 97.81 \\ 
	Trains                      & \textbf{99.9}  & 99.84 & \textbf{100.0}   & 99.46 & \textbf{100.0}  & 99.98 & 99.97 & 99.74 & 99.97 \\ 
	Stadium seats               & 99.92 & \textbf{100.0} & \textbf{100.0}   & 99.98 & 99.89 & 99.99 & 99.98 & 99.99 & \textbf{100.0}   \\ \hline
	OA                          & 86.95 & 89.09 & 90.05 & 91.02 & 91.24 & 89.98 & 90.56 & 90.31 & \textbf{92.38} \\ 
	AA                          & 92.36 & 91.91 & 93.14 & 95.24 & 95.02 & 94.02 & 94.74 & 94.61 & \textbf{95.51} \\ 
	Kappa                       & 83.39 & 85.96 & 87.2  & 88.46 & 88.71 & 87.14 & 87.81 & 87.55 & \textbf{90.16} \\ 
\bottomrule
\end{tabular}}
\label{houston2018_compare_table}
\end{table*}

\begin{figure*}[h!]
\centering
\includegraphics [width=0.95\textwidth]{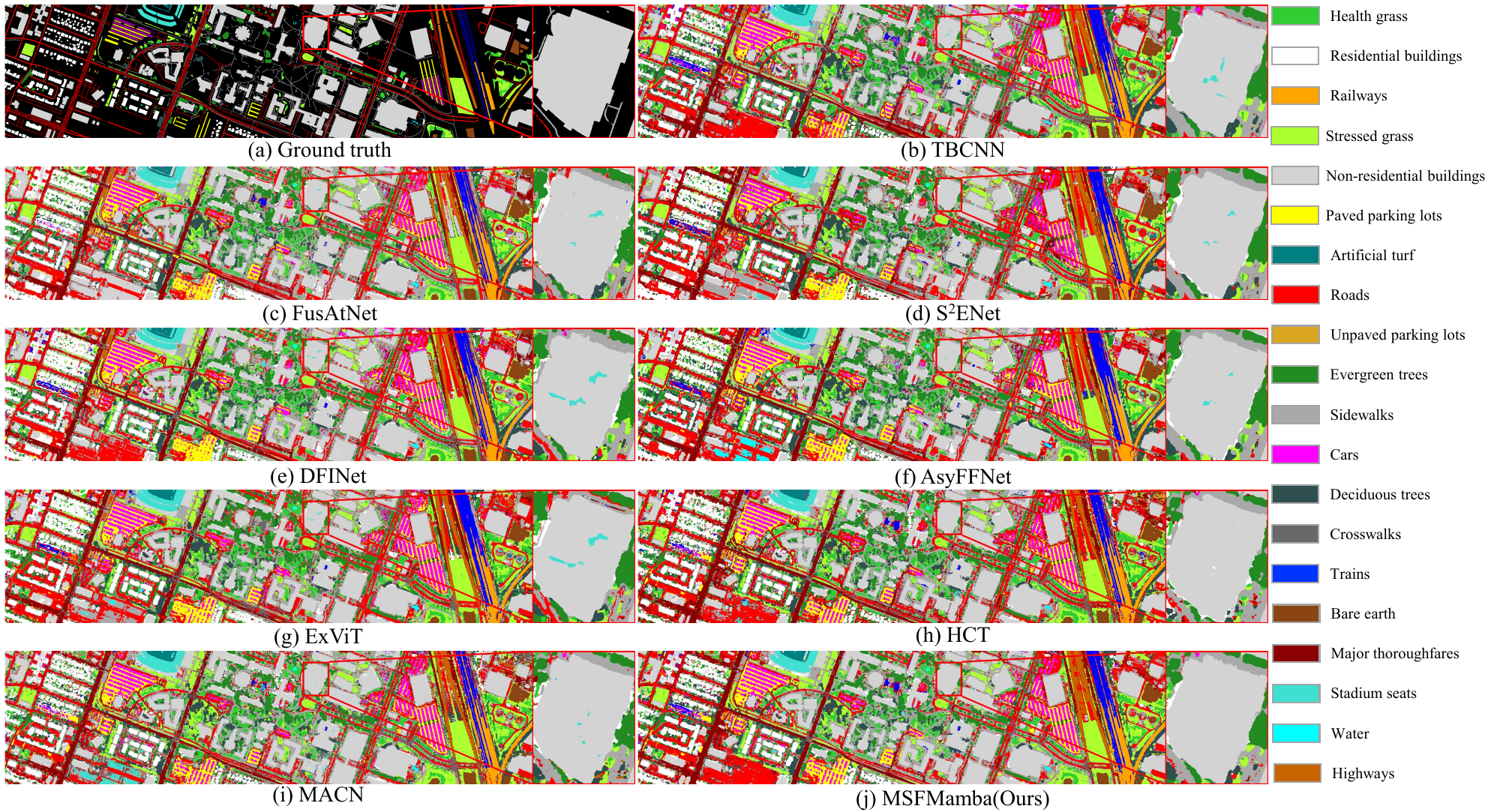}
\caption{Classification results of different methods for the Houston 2018 dataset.}
\label{houston2018_fig_com}
\end{figure*}

\begin{table*}[h!]
\centering
\renewcommand\arraystretch{1.1}
\caption{{Classification performance of different methods on the Houston 2013 dataset.}}
\scalebox{0.9}{
\begin{tabular}{c|ccccccccc}
\hline
Class           & TBCNN  & FusAtNet & $S^2$ENet & DFINet & AsyFFNet & ExViT & HCT    & MACN   & MSFMamba \\ \hline
Health grass    & 82.72  & 82.53    & 81.48     & \textbf82.91  & 82.72    & 82.72  & 81.1   & 82.90  & 82.72    \\
Stressed grass  & 83.18  & 85.15    & 84.68     & 85.15  & 83.93    & 85.15  & 86.09  & \textbf{100.0}    & 99.44    \\
Synthetic grass & \textbf{100.0} & 98.81    & 99.01     & 99.60  & \textbf{100.0}   & 99.21  & 99.80  & 99.80  & 98.81    \\
Trees           & 94.03  & 92.90    & 92.14     & 92.33  & 91.95    & 91.95  & 98.01  & 98.11  & \textbf{98.39}    \\
Soil            & 99.15  & \textbf{100.0}   & \textbf{100.0}    & 99.81  & \textbf{100.0}   & \textbf{100.0} & \textbf{100.0} & \textbf{100.0} & \textbf{100.0}   \\
Water           & 99.30  & 98.60    & \textbf{100.0}    & \textbf{100.0} & 95.80    & 98.60  & 95.10  & 95.80  & 95.80    \\
Residential     & 79.66  & 85.45    & 89.93     & 93.47  & \textbf{96.18}    & 86.66  & 83.49  & 78.17  & 82.18    \\
Commercial      & 55.27  & 81.81    & 90.18     & 82.58  & 81.14    & 91.05  & 86.43  & 80.46  & \textbf{94.23}    \\
Road            & 75.83  & 83.76    & 92.63     & 89.05  & 86.12    & 91.12  & 86.69  & 90.93  & \textbf{93.96}    \\
Highway         & 62.55  & 53.67    & 64.86     & 56.56  & 64.58    & 64.86  & 79.05  & 55.88  & \textbf{79.92}    \\
Railway         & 96.87  & 79.65    & 97.60     & 94.15  & 86.76    & 77.06  & 95.97  & \textbf{99.33}  & 93.47    \\
Parking lot 1   & 86.55  & 91.26    & 88.09     & 94.72  & 90.59    & 88.76  & 98.46  & \textbf{99.71}  & 96.73    \\
Parking lot 2   & 53.68  & 81.05    & 91.93     & 89.12  & \textbf{92.28}    & 87.72  & 90.88  & 88.77  & 90.88    \\
Tennis court    & 98.79  & \textbf{100.0}   & \textbf{100.0}    & \textbf{100.0} & 95.14    & 97.17  & \textbf{100.0} & \textbf{100.0} & \textbf{100.0}   \\
Running track   & 98.10  & 98.73    & \textbf100.0    & \textbf100.0 & \textbf100.0   & 98.52  & \textbf100.0 & \textbf100.0 & \textbf100.0   \\ \hline
OA              & 82.91  & 85.32    & 89.56     & 88.59  & 87.95    & 87.42  & 90.66  & 89.81  & \textbf92.86    \\
AA              & 84.38  & 87.56    & 91.50     & 90.63  & 89.81    & 89.37  & 92.07  & 91.33  & \textbf{93.77}    \\
Kappa           & 81.43  & 84.12    & 88.69     & 87.64  & 86.96    & 86.38  & 89.86  & 88.93  & \textbf{92.25}    \\ \hline
\end{tabular}}
\label{houston2013_compare_table}
\end{table*}

\begin{figure*}[h!]
\centering
\includegraphics [width=0.95\textwidth]{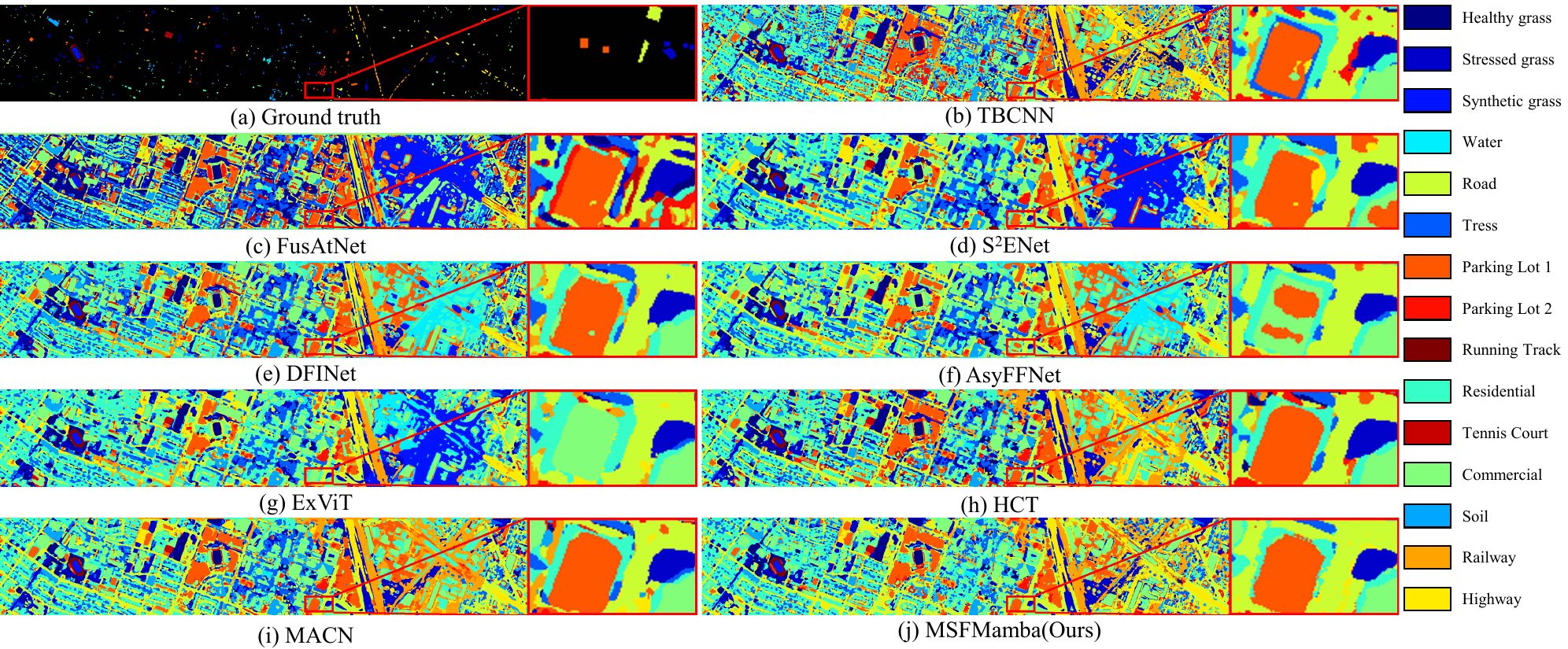}
\caption{Classification results of different methods for the Houston 2013 dataset.}
\label{houston2013_fig_com}
\end{figure*}

\subsection{Performance Comparison}

We compared the proposed MSFMamba with existing state-of-the-art methods, including Two-Branch CNN(TBCNN) \cite{TBCNN}, FusAtNet,Spatial-Spectral Cross-Modal Enhancement Network($S^2$ENet) \cite{S$^2$ENet}, Depthwise Feature Interaction Network(DFINet) \cite{DFINet}, Asymmetric Feature Fusion Network(AsyFFNet) \cite{AsyFFNet}, ExViT, Hierarchical CNN and Transformer(HCT) \cite{HCT} and Mixing self-Attention and Convolution Network(MACN) \cite{li2023mixing}. 
TBCNN \cite{TBCNN} investigates the classification fusion of hyperspectral imagery and data from other sensors using a two-branch convolution neural network. FusAtNet \cite{FusAtNet} presents a multi-source classification framework by utilizing a self-attention mechanism for spectral features and a cross-attention approach for spatial features. 
$S^2$ENet \cite{S$^2$ENet} proposes a spatial-spectral enhancement module for cross-modal information interaction, which effectively facilitates the feature interaction between multi-source data. DFINet \cite{DFINet} uses a depth-wise cross attention module to extract complementary information from multisource feature pairs.  AsyFFNet \cite{AsyFFNet} is a multi-source data classification method based on asymmetric feature fusion, which employs weight-share residual blocks for feature extraction and a feature calibration module for the spatial-wise multi-source feature modeling. ExViT \cite{ExVit} utilizes parallel branches of position-shared Transformer extended with separable convolution modules to process multi-modal image patches.  In MACN \cite{li2023mixing}, mixing self-attention and convolution Transformer layer is proposed to extract local and global multi-scale feature perception. The classification performance of various methods on multiple datasets is presented in Tables \ref{berlin_compare_table}, \ref{augsburg_compare_table}, \ref{houston2018_compare_table}, and \ref{houston2013_compare_table}. These tables compare the recall rates of different methods for specific classes, providing a detailed evaluation of their performance.

\begin{table*}[h]
\centering
\caption{Influence of MSpa-Mamba, Spe-Mamba and Fus-Mamba on classification results of MSFMamba.}
\renewcommand{\arraystretch}{1.1}
\scalebox{0.9}{
\begin{tabular}{ccc|cccc} \hline
\toprule
MSpa-Mamba &  Spe-Mamba & ~~~ Fus-Mamba ~~~ & ~~~~ Berlin ~~~~ & Augsburg & Houston2018 & Houston2013\\
\midrule
\faCheck & {\color{gray}\faTimes} & {\color{gray}\faTimes} & 74.88 & 89.05 & 89.56 & 90.27\\ 
\faCheck & \faCheck & {\color{gray}\faTimes} & 76.00 & 90.60 & 91.44 & 91.55\\ 
\faCheck & {\color{gray}\faTimes} & \faCheck & 75.88 & 90.55 & 91.76 & 91.82\\ 
\faCheck & \faCheck & \faCheck & 76.92 & 91.38 & 92.38 & 92.86\\   
\bottomrule \hline
\end{tabular}}
\label{table_ablation_compare}
\end{table*}

\textit{1) Results on the Berlin Dataset.}  
The classification performance of various methods on the Berlin dataset is presented in Table \ref{berlin_compare_table}, with corresponding classification maps shown in Fig. \ref{berlin_fig_com}.
Our proposed MSFMamba achieves the highest OA of 76.92\%, an AA of 64.88\%, and a Kappa coefficient of 64.88\%. These results highlight the model’s robust multi-scale spatial contextual modeling capabilities, enabled by its sequential modeling approach.
A closer analysis reveals that MSFMamba excels in distinguishing fine-grained classes, particularly Industrial Area, Low Plants, and Allotment, where capturing intricate textures and boundary details is critical. The model effectively handles these challenges, resulting in more accurate classification. In the visualized results, particularly in the magnified sections, it is evident that MSFMamba produces classification maps much closer to the ground truth. While other methods tend to misclassify areas like Commercial Area, MSFMamba more accurately captures these regions, showcasing its ability to preserve finer details and textures in challenging areas.

\textit{2) Results on the Augsburg Dataset.} 
Table \ref{augsburg_compare_table} presents the quantitative results of various methods on the Augsburg dataset. A closer analysis of class-wise performance reveals that MSFMamba achieves notable improvements in categories such as Low Plants, Allotment and Water, which are characterized by complex spectral-spatial patterns. Our MSFMamba outperforms the second-best algorithm by 0.22\% in OA, 0.12\% in AA, and 0.39\% in Kappa. The corresponding classification maps, shown in Fig. \ref{augsburg_fig_com}, further demonstrate the advantage of MSFMamba. By effectively extracting spatial-spectral semantic features through its sequential modeling mechanism, MSFMamba produces clearer boundaries and significantly reduces misclassified regions compared to other methods.

\textit{3) Results on the Houston2018 Dataset.} 
Table \ref{houston2018_compare_table} presents the classification performance of various methods on the Houston2018 dataset, with Fig. \ref{houston2018_fig_com} showing the corresponding classification maps. MSFMamba delivers a highly refined classification map, with most noisy regions significantly reduced, resulting in cleaner and more accurate segmentation. On the Houston2018 dataset, MSFMamba demonstrates exceptional class-specific performance, especially in categories like Stressed grass, Evergreen trees, Non-residential buildings and Roads, where it surpasses other methods by better capturing spatial structures and avoiding common misclassifications. The sequential scanning approach employed by MSFMamba proves to be highly effective for HSI and LiDAR data fusion, leading to more accurate and consistent class predictions, as evidenced by the minimal noise and misclassified areas in the visualized results.

\textit{4) Results on the Houston2013 Dataset.}
Table \ref{houston2013_compare_table} summarizes the classification results of various approaches on the Houston2013 dataset, while Fig. \ref{houston2013_fig_com} illustrates the respective classification maps. MSFMamba produces a well-refined classification output, achieving more precise segmentation and significantly minimizing noise in most areas, resulting in a clearer result overall. On the Houston2013 dataset, MSFMamba exhibits outstanding performance on specific classes, particularly in categories like Trees, Commercial, Road and Highway. It outperforms other methods by accurately capturing spatial structures and mitigating common misclassification errors.

\textit{5) Practical Implications.} 
The experimental results across all datasets highlight MSFMamba's applicability in addressing real-world challenges such as land cover classification and urban infrastructure planning. For example, its high classification accuracy on datasets like Berlin and Augsburg makes it valuable for urban planning and ecological monitoring, while its robust performance on Houston2018 and Houston2013 datasets supports transportation network analysis and city development projects.

\begin{figure*}[ht]
\centering
\subfigure[]{\includegraphics[width=.15\textwidth]{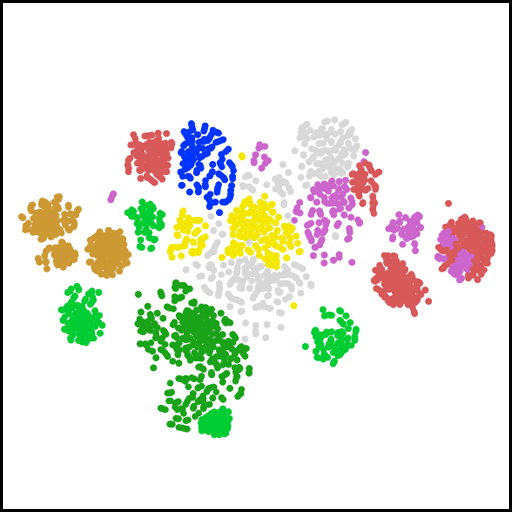}}
\subfigure[]{\includegraphics[width=.15\textwidth]{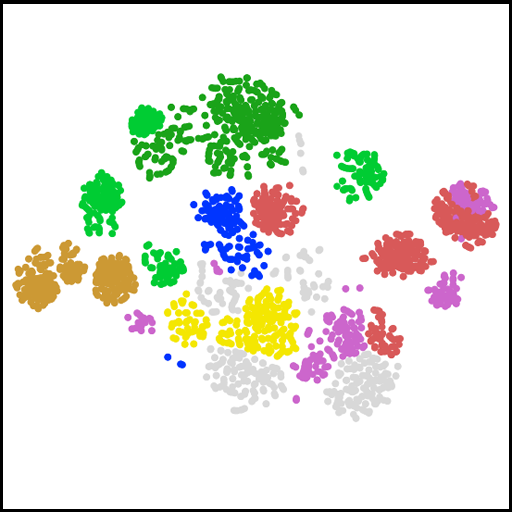}}
\subfigure[]{\includegraphics[width=.15\textwidth]{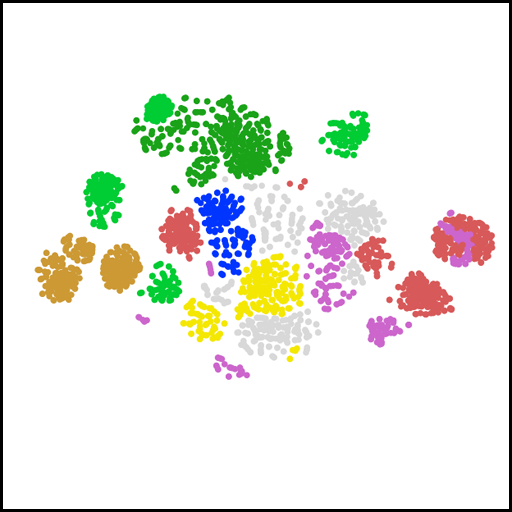}}
\subfigure[]{\includegraphics[width=.15\textwidth]{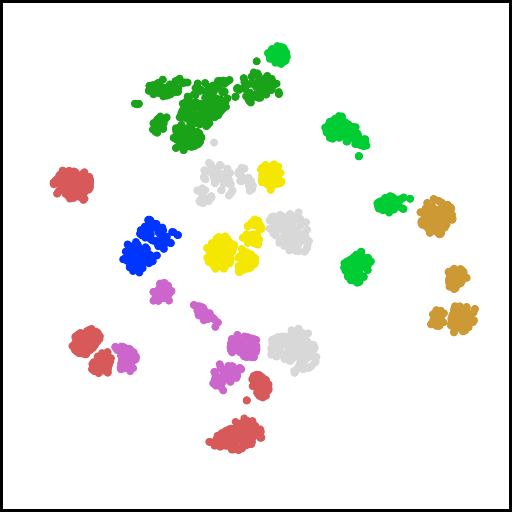}}
\caption{t-SNE visualization of MSFMamba with different components. (a) MSpa-Mamba. (b) MSpa-Mamba + Spe-Mamba. (c) MSpa-Mamba + Fus-Mamba. (d) Full model of our MSFMamba.}
\label{fig_tsne}
\end{figure*}

\subsection{Ablation Study}

Table \ref{table_ablation_compare} presents the classification performance of different module combinations on the four datasets. The results reveal a clear trend: the inclusion of each module consistently improves the OA. MSpa-Mamba serves as the fundamental module, achieving OA values of 74.88\%, 89.05\%, 89.56\%, and 90.27\% on the four datasets, respectively. Its strong spatial feature extraction capability lays a solid foundation for classification tasks. Adding Spe-Mamba further enhances performance to 76.00\%, 90.60\%, 91.44\%, and 91.55\%. This significant improvement underscores the importance of spectral feature extraction in capturing the unique patterns of HSI and SAR/LiDAR data. The combination of MSpa-Mamba and Fus-Mamba yields accuracies of 75.88\%, 90.55\%, 91.76\%, and 91.82\%. This demonstrates the critical role of effective multi-source data fusion in handling heterogeneous datasets. When all three components are combined, the model achieves the best performance across four datasets, with substantial noise reduction and more precise class boundaries. This demonstrates the complementary nature of the modules: MSpa-Mamba and Spe-Mamba provide robust spatial-spectral feature extraction, while Fus-Mamba effectively integrates these features, maximizing classification accuracy.

We conducted t-SNE (t-distributed Stochastic Neighbor Embedding) visualizations of the three components on the Berlin dataset for our MSFMamba, and the results are illustrated in Fig. \ref{fig_tsne}. t-SNE is a dimensionality reduction technique widely used to visualize high-dimensional data by projecting it into a lower-dimensional space while preserving its local structure. It can be observed that with only the MSpa-Mamba, the features of different classes are less compact. With all the three modules, the feature representations display the most distinct and well-defined clusters compared to the others. It is evident that the sequential scanning model generate more robust feature representations for multi-source data classification.

\begin{table}[h!]
        \caption{Experimental Results of the Multi-Scan Strategy on the Berlin Dataset.}
	\centering
      \renewcommand{\arraystretch}{1.3}
      \scalebox{0.9}{
	\begin{tabular}{cccc}
		\hline\toprule
		~~ \makecell[c]{Path number for \\downsampling} ~~ & OA  & AA   & Kappa \\ \hline
		0          & 74.71 & 61.15 & 61.99 \\
		1          & 75.44 & 62.85 & 62.93 \\
		2          & 76.92 & 64.88 & 64.88 \\
		3          & 75.30 & 62.50 & 62.91 \\
		4          & 74.31 & 61.79 & 61.99 \\ 
      \bottomrule\hline
	\end{tabular}}
	\label{multi-scale analysis}
\end{table}
\subsection{Analysis of Multi-Scale Spatial Mamba}

Table \ref{multi-scale analysis} presents the impact of varying the number of downsampling paths on the classification performance for the Berlin dataset. The results show that the best performance is achieved when two paths retain the original resolution and two paths are downsampled, yielding OA, AA, and Kappa values of 76.92\%, 64.88\%, and 64.88\%, respectively. This configuration balances the reduction of redundant features with the preservation of critical details, demonstrating the effectiveness of the multi-scale scanning approach. In contrast, configurations with three or more downsampling paths lead to a performance decline, with OA dropping to 75.30\% and 74.31\%, highlighting the trade-off between feature reduction and information loss.

{This analysis emphasizes the importance of integrating original-resolution and downsampled paths to capture both fine-grained details and broader spatial patterns. The consistent improvement in classification metrics from 0 to 2 downsampling paths underscores the contribution of this multi-scale strategy in handling complex spectral-spatial patterns in the Berlin dataset. These results demonstrate that an optimal balance in downsampling is critical for achieving robust performance in remote sensing classification tasks.}

\begin{figure}[ht!]
\centering
\subfigure[]{\includegraphics[width=1.4in]{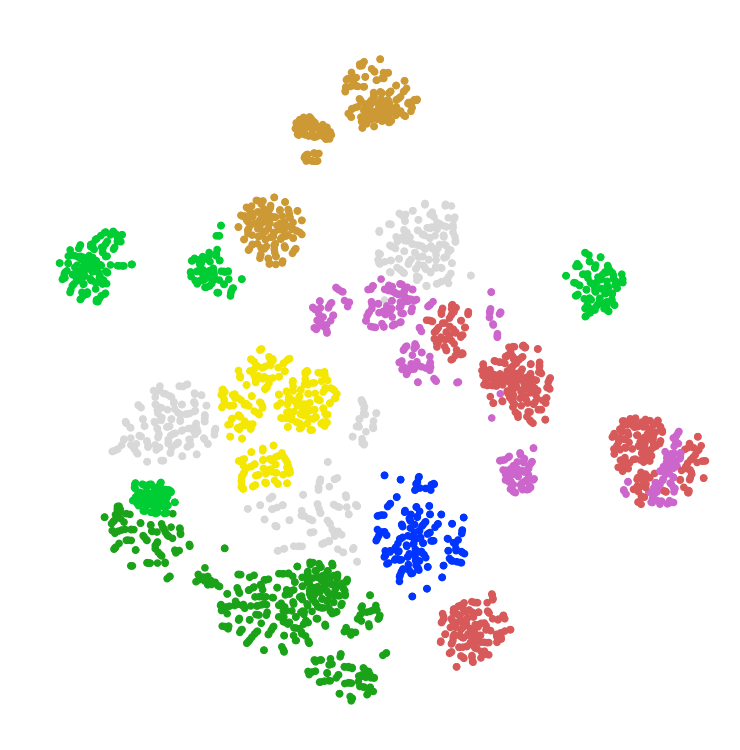}}
\subfigure[]{\includegraphics[width=1.4in]{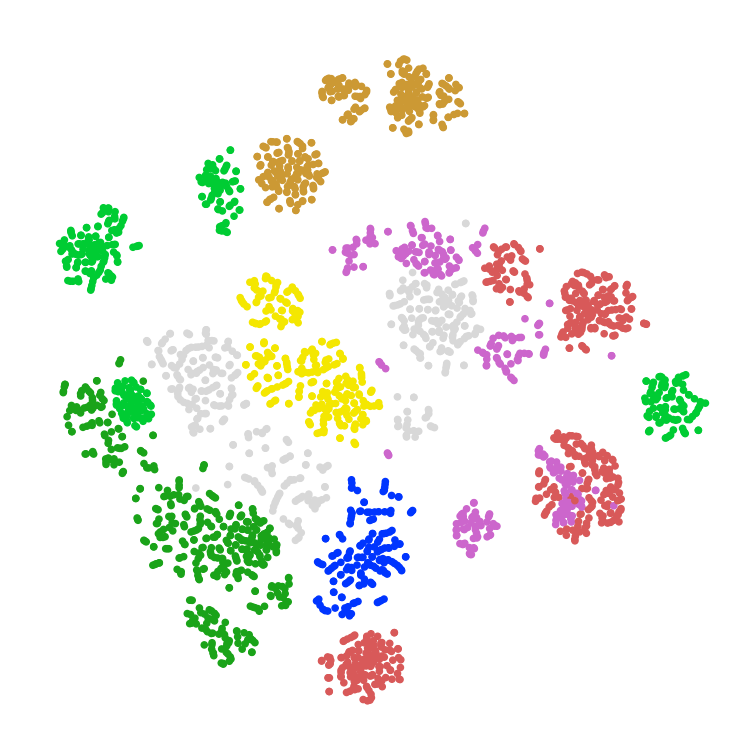}}
\newline
\subfigure[]{\includegraphics[width=1.4in]{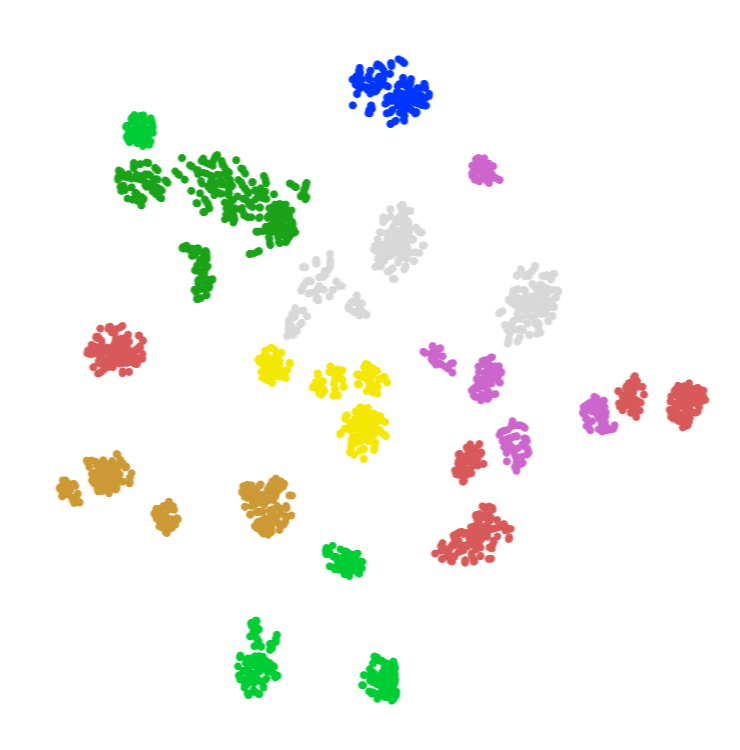}}
\subfigure[]{\includegraphics[width=1.4in]{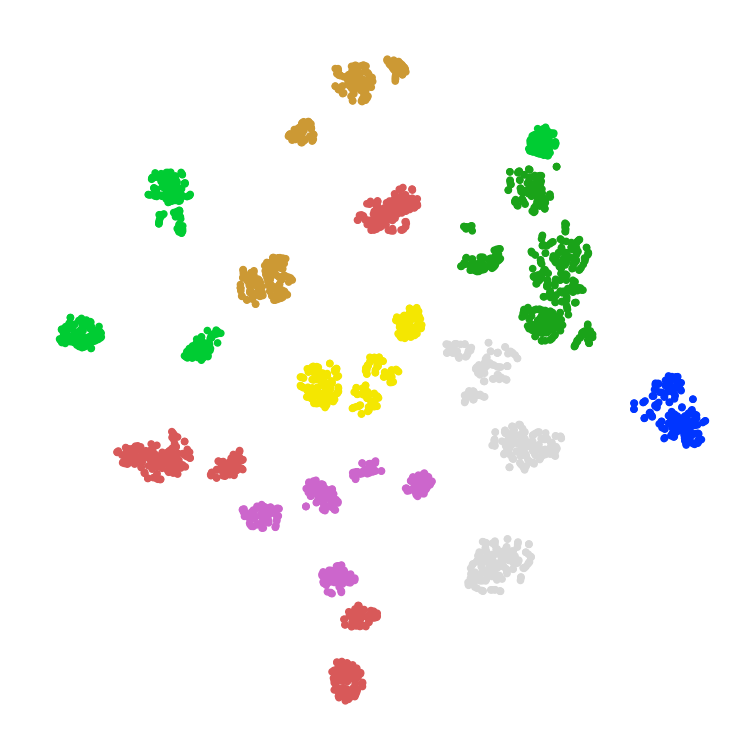}}
\caption{t-SNE Visualization of Original and Fused HSI-SAR Features in Fusion Mamba.(a) Original HSI Features. 
(b) Original SAR Features. (c) Fused HSI Features after Fusion Mamba. (d) Fused SAR Features after Fusion Mamba.}
\label{fig_tsne_fuse}
\end{figure}

\begin{table}[ht!]
\centering
\caption{Impact of Scanning Routes on Classification Performance}
\begin{tabular}{c|ccc}
\hline\toprule
Scanning Routes & OA(\%) & AA(\%) & Kappa \\
\midrule
2 & 75.28 & 63.22 & 62.28 \\
4 & 76.92 & 64.88 & 64.88 \\
6 & 74.98 & 63.45 & 63.39 \\
\bottomrule\hline
\end{tabular}
\label{Scanning Routes}
\end{table}

\begin{table*}[h!]
\centering
\caption{Comparative Analysis of Model Parameters, FLOPS and Inference time on the Augsburg Dataset}
\renewcommand{\arraystretch}{1.1}
\scalebox{0.9}{
\begin{tabular}{c|cccccccccc}
\hline\toprule
Metrics  & FusAtNet & $S^2$ENet & DFINet & AsyFFNet & ExVit  & HCT    & MACN   & MSFMamba &  \\ 
\midrule
Params (M) & 37.7177  & 1.4701    & 1.3155 & 2.4132   & 1.8848 & 4.7811 & 2.0199 & 1.5252   &  \\
FLOPS (G)  & 3.5619   & 0.1779    & 0.1191 & 0.2064   & 0.2901 & 0.0437 & 0.1672 & 0.0377  &  \\
Inference time (s) & 0.2469  & 0.2489  & 0.3182 & 0.2924 & 0.3271 & 0.7524 & 0.1917 & 0.1747  &\\
\bottomrule\hline
\end{tabular}}
\label{table_complex}
\end{table*}

\subsection{Analysis of Fusion Mamba Effects}

We aim to understand how the Fusion Mamba block enhances feature representation in multi-source remote sensing data by using t-SNE visualizations. The comparison of features before and after fusion provides insights into the strengths of the Fusion Mamba block in integrating diverse data modalities.

Fig. \ref{fig_tsne_fuse} presents the t-SNE visualizations of HSI and SAR features before and after applying the Fusion Mamba block:
Fig. \ref{fig_tsne_fuse}(a) shows the original HSI features before fusion.
Fig. \ref{fig_tsne_fuse}(b) presents the original SAR features before fusion.
In both Fig. \ref{fig_tsne_fuse}(a) and Fig. \ref{fig_tsne_fuse}(b), the separability between different classes is relatively poor, with noticeable overlap between clusters. This suggests that the original features alone struggle to clearly discriminate between certain categories.

Fig. \ref{fig_tsne_fuse}(c) demonstrates the fused HSI features after the Fusion Mamba block is applied.
Fig. \ref{fig_tsne_fuse}(d) displays the fused SAR features after the fusion process.
The visualizations in Fig. \ref{fig_tsne_fuse}(c) and Fig. \ref{fig_tsne_fuse}(d) show a significant improvement in class distinction, with clearer and more compact clusters. This indicates that the Fusion Mamba block successfully integrates features from different modalities, enhancing the overall representation and improving the discriminative power of the fused features.

\subsection{Analysis of Scanning Route Numbers}

The number of scanning routes is an important parameter in our MSFMamba model, as it directly impacts the diversity of spatial feature representations. In this section, we evaluate the effect of different numbers of scanning routes on classification performance and computational efficiency. Specifically, we compared configurations with 2, 4, and 6 scanning routes using the Berlin dataset. In the case of 2 routes, row-wise and column-wise scanning is utilized. In the case of 4 routes, row-wise, column-wise, reverse row-wise, and reverse column wise scanning is used. To achieve 6 routes, two diagonal scanning routes (top-left to bottom right, and top-right to bottom-left) are added. 

From Table \ref{Scanning Routes}, we observe that the classification performance improves with the addition of reverse scanning directions. The model with 4 scanning routes achieved the highest performance, indicating that adding more scanning routes enhances the spatial feature diversity and improves the model's ability to differentiate between classes. However, the configuration with 6 scanning routes, while providing more feature diversity, results in slightly lower performance. This could be due to the increased complexity, which may introduce more noise into the features or lead to overfitting on certain classes.
Our analysis shows that using 4 scanning routes offers the best classification performance in terms of accuracy. These results suggest that adding more scanning routes does not always result in a proportional improvement in performance and may even lead to diminishing returns.

\subsection{Analysis of Computational Complexity}

To provide a comprehensive computational complexity analysis, we evaluated the model parameters, floating point operations per second(FLOPS), and inference time of various methods on the Augsburg dataset, as shown in Table \ref{table_complex}. Specifically, we measure the total number of learnable parameters in millions (M), the computational cost in gigaflops (G), and the inference time in seconds (s). Our proposed MSFMamba model exhibits the lowest FLOPs and inference time, confirming its computational efficiency compared to other methods. This advantage highlights MSFMamba's suitability for real-time applications, especially on resource-constrained edge devices like drones and mobile platforms. Furthermore, MSFMamba requires significantly fewer parameters than most state-of-the-art methods, contributing to reduced memory requirements and facilitating efficient deployment.

\section{Conclusions and Future Work}

This paper introduced MSFMamba, a novel network designed for the joint classification of multi-source remote sensing data. The network utilized SSM-based blocks to achieve linear complexity and a large receptive field. The MSpa-Mamba block efficiently extracted spatial features using a multi-scale strategy, reducing computational and feature redundancies. The Spe-Mamba block enhanced spectral feature extraction, enabling effective HSI and LiDAR/SAR fusion, while the Fus-Mamba block facilitated robust cross-modal feature integration by extending the original Mamba design to support dual inputs. Experiments on four datasets demonstrated that MSFMamba achieved state-of-the-art performance, highlighting its robustness, efficiency, and potential for real-world applications.

However, the model has certain limitations, such as reduced accuracy for imbalanced classes and limited interpretability in critical applications like urban planning and disaster management. Addressing these challenges requires exploring techniques such as dynamic sampling, advanced loss functions, and methods to enhance explainability. 
Future work will focus on developing hybrid models that combine Mamba with Transformer architectures to further improve long-range feature modeling and multi-modal data fusion. Additionally, methods for improving the interpretability of MSFMamba will be explored, particularly in terms of how it processes and integrates the contributions of different data modalities.Investigating feature importance and feature selection techniques will also be part of the future research, ensuring that the model is both efficient and transparent in its operations.

\bibliography{source}

\begin{thebibliography}{10}
\providecommand{\url}[1]{#1}
\csname url@samestyle\endcsname
\providecommand{\newblock}{\relax}
\providecommand{\bibinfo}[2]{#2}
\providecommand{\BIBentrySTDinterwordspacing}{\spaceskip=0pt\relax}
\providecommand{\BIBentryALTinterwordstretchfactor}{4}
\providecommand{\BIBentryALTinterwordspacing}{\spaceskip=\fontdimen2\font plus
\BIBentryALTinterwordstretchfactor\fontdimen3\font minus \fontdimen4\font\relax}
\providecommand{\BIBforeignlanguage}[2]{{%
\expandafter\ifx\csname l@#1\endcsname\relax
\typeout{** WARNING: IEEEtran.bst: No hyphenation pattern has been}%
\typeout{** loaded for the language `#1'. Using the pattern for}%
\typeout{** the default language instead.}%
\else
\language=\csname l@#1\endcsname
\fi
#2}}
\providecommand{\BIBdecl}{\relax}
\BIBdecl

\bibitem{zhmm17jars}
M.~Zhang, W.~Li, and Q.~Du, ``Collaborative classification of hyperspectral and visible images with convolutional neural network,'' \emph{Journal of Applied Remote Sensing}, vol.~11, pp. 1--12, 09 2017.

\bibitem{weixin_join1}
M.~Mohammadi and A.~Sharifi, ``Evaluation of convolutional neural networks for urban mapping using satellite images,'' \emph{Journal of the Indian Society of Remote Sensing}, vol.~49, no.~9, pp. 2125--2131, 2021.

\bibitem{weixin_join2}
A.~Sharifi and S.~Felegari, ``Nitrogen dioxide (no2) pollution monitoring with sentinel-5p satellite imagery over during the coronavirus pandemic (case study: Tehran),'' \emph{Remote Sensing Letters}, vol.~13, no.~10, pp. 1029--1039, 2022.

\bibitem{Latent_Reconstruction}
J.~Yue, L.~Fang, and M.~He, ``Spectral-spatial latent reconstruction for open-set hyperspectral image classification,'' \emph{IEEE Transactions on Image Processing}, vol.~31, pp. 5227--5241, 2022.

\bibitem{mineral_exploration}
H.~Shirmard, E.~Farahbakhsh, R.~D. Müller, and R.~Chandra, ``A review of machine learning in processing remote sensing data for mineral exploration,'' \emph{Remote Sensing of Environment}, vol. 268, pp. 1--21, 2022.

\bibitem{dynamic_Super-Pixel}
C.~Wang, L.~Zhang, W.~Wei, and Y.~Zhang, ``Dynamic super-pixel normalization for robust hyperspectral image classification,'' \emph{IEEE Transactions on Geoscience and Remote Sensing}, vol.~61, pp. 1--13, 2023.

\bibitem{Terrain}
N.~Jiang, H.-B. Li, C.-J. Li, H.-X. Xiao, and J.-W. Zhou, ``A fusion method using terrestrial laser scanning and unmanned aerial vehicle photogrammetry for landslide deformation monitoring under complex terrain conditions,'' \emph{IEEE Transactions on Geoscience and Remote Sensing}, vol.~60, pp. 1--14, 2022.

\bibitem{agriculture}
S.~Khanal, K.~Kc, J.~P. Fulton, S.~Shearer, and E.~Ozkan, ``Remote sensing in agriculture—accomplishments, limitations, and opportunities,'' \emph{Remote Sensing}, vol.~12, no.~22, p. 3783, 2020.

\bibitem{ganhan_yaogan_join1}
S.~Jalayer, A.~Sharifi, D.~Abbasi-Moghadam, A.~Tariq, and S.~Qin, ``Assessment of spatiotemporal characteristic of droughts using in situ and remote sensing-based drought indices,'' \emph{IEEE Journal of Selected Topics in Applied Earth Observations and Remote Sensing}, vol.~16, pp. 1483--1502, 2023.

\bibitem{hzy24jstars}
Z.~Huang, J.~Cheng, G.~Wei, X.~Hua, and Y.~Wang, ``An urban land cover classification method based on segments’ multidimension feature fusion,'' \emph{IEEE Journal of Selected Topics in Applied Earth Observations and Remote Sensing}, vol.~17, pp. 5580--5593, 2024.

\bibitem{kk15jstars}
K.~Karantzalos, D.~Bliziotis, and A.~Karmas, ``A scalable geospatial web service for near real-time, high-resolution land cover mapping,'' \emph{IEEE Journal of Selected Topics in Applied Earth Observations and Remote Sensing}, vol.~8, no.~10, pp. 4665--4674, 2015.

\bibitem{LiDAR_join1}
S.~Yu, D.~Guan, Z.~Gu, J.~Guo, Z.~Liu, and Y.~Liu, ``Radar target complex high-resolution range profile modulation by external time coding metasurface,'' \emph{IEEE Transactions on Microwave Theory and Techniques}, 2024.

\bibitem{lz24tgrs}
Z.~Li, K.~Zheng, J.~Li, C.~Li, and L.~Gao, ``Cross-semantic heterogeneous modeling network for hyperspectral image classification,'' \emph{IEEE Transactions on Geoscience and Remote Sensing}, vol.~62, pp. 1--16, 2024.

\bibitem{zx24grsl}
X.~Zhao, M.~Song, and T.~Yang, ``Hyperspectral unmixing based on chaotic sequence optimization of lp norm,'' \emph{IEEE Geoscience and Remote Sensing Letters}, vol.~21, pp. 1--5, 2024.

\bibitem{hdf21grsm}
D.~Hong, W.~He, N.~Yokoya, J.~Yao, L.~Gao, L.~Zhang, J.~Chanussot, and X.~Zhu, ``Interpretable hyperspectral artificial intelligence: When nonconvex modeling meets hyperspectral remote sensing,'' \emph{IEEE Geoscience and Remote Sensing Magazine}, vol.~9, no.~2, pp. 52--87, 2021.

\bibitem{Range_Resolved}
P.~W. Milonni, ``Lidar. range-resolved optical remote sensing of the atmosphere, in the springer series in optical sciences, vol. 102, edited by claus weitkamp: Scope: review. level: specialist,'' 2009.

\bibitem{Terrestrial_Lidar}
M.~Pashaei, M.~J. Starek, and J.~Berryhill, ``Full-waveform terrestrial lidar data classification using raw digitized waveform signals,'' in \emph{IEEE International Geoscience and Remote Sensing Symposium (IGARSS)}, 2022, pp. 1916--1919.

\bibitem{polarimetric_SAR}
J.~Shi and H.~Jin, ``Riemannian nearest-regularized subspace classification for polarimetric sar images,'' \emph{IEEE Geoscience and Remote Sensing Letters}, vol.~19, pp. 1--5, 2022.

\bibitem{time-series_PolSAR}
J.~Cheng, D.~Xiang, Q.~Yin, and F.~Zhang, ``A novel crop classification method based on the tensor-gcn for time-series polsar data,'' \emph{IEEE Transactions on Geoscience and Remote Sensing}, vol.~60, pp. 1--14, 2022.

\bibitem{Multitemporal_SAR}
J.~Ni, C.~L{\'o}pez-Mart{\'\i}nez, Z.~Hu, and F.~Zhang, ``Multitemporal sar and polarimetric sar optimization and classification: Reinterpreting temporal coherence,'' \emph{IEEE Transactions on Geoscience and Remote Sensing}, vol.~60, pp. 1--17, 2022.

\bibitem{yj24tgrs}
J.~X. Yang, J.~Zhou, J.~Wang, H.~Tian, and A.~W.-C. Liew, ``Lidar-guided cross-attention fusion for hyperspectral band selection and image classification,'' \emph{IEEE Transactions on Geoscience and Remote Sensing}, vol.~62, pp. 1--15, 2024.

\bibitem{mxc24tgrs}
X.~Meng, S.~Zhang, Q.~Liu, G.~Yang, and W.~Sun, ``Uncertain category-aware fusion network for hyperspectral and lidar joint classification,'' \emph{IEEE Transactions on Geoscience and Remote Sensing}, vol.~62, pp. 1--15, 2024.

\bibitem{gyh24tip}
Y.~Gao, W.~Li, J.~Wang, M.~Zhang, and R.~Tao, ``Relationship learning from multisource images via spatial-spectral perception network,'' \emph{IEEE Transactions on Image Processing}, vol.~33, pp. 3271--3284, 2024.

\bibitem{fusion_join2}
J.~Chen, Q.~Wang, W.~Peng, H.~Xu, X.~Li, and W.~Xu, ``Disparity-based multiscale fusion network for transportation detection,'' \emph{IEEE Transactions on Intelligent Transportation Systems}, vol.~23, no.~10, pp. 18\,855--18\,863, 2022.

\bibitem{zmt16grsl}
M.~Zheng and Y.~Zhang, ``Dem-aided bundle adjustment with multisource satellite imagery: Zy-3 and gf-1 in large areas,'' \emph{IEEE Geoscience and Remote Sensing Letters}, vol.~13, no.~6, pp. 880--884, 2016.

\bibitem{xzg18grsl}
Z.~Xu, J.~Chen, J.~Xia, P.~Du, H.~Zheng, and L.~Gan, ``Multisource earth observation data for land-cover classification using random forest,'' \emph{IEEE Geoscience and Remote Sensing Letters}, vol.~15, no.~5, pp. 789--793, 2018.

\bibitem{xys24tgrs}
Y.~Zhang, S.~Yan, X.~Jiang, L.~Zhang, Z.~Cai, and J.~Li, ``Dual graph learning affinity propagation for multimodal remote sensing image clustering,'' \emph{IEEE Transactions on Geoscience and Remote Sensing}, vol.~62, pp. 1--13, 2024.

\bibitem{lzy23grsl}
Z.~Lv, H.~Huang, W.~Sun, T.~Lei, J.~A. Benediktsson, and J.~Li, ``Novel enhanced unet for change detection using multimodal remote sensing image,'' \emph{IEEE Geoscience and Remote Sensing Letters}, vol.~20, pp. 1--5, 2023.

\bibitem{Indian}
M.~K. Singh, S.~Mohan, and B.~Kumar, ``Fusion of hyperspectral and lidar data using sparse stacked autoencoder for land cover classification with 3d-2d convolutional neural network,'' \emph{Journal of Applied Remote Sensing}, vol.~16, no.~3, pp. 034\,523--034\,523, 2022.

\bibitem{DMSCA}
H.~W, Yu~and F, ``Dmsca: deep multiscale cross-modal attention network for hyperspectral and light detection and ranging data fusion and joint classification,'' \emph{Journal of Applied Remote Sensing}, vol.~18, no.~3, pp. 036\,505--036\,505, 2024.

\bibitem{dwq22tgrs}
W.~Dong, T.~Zhang, J.~Qu, S.~Xiao, T.~Zhang, and Y.~Li, ``Multibranch feature fusion network with self- and cross-guided attention for hyperspectral and lidar classification,'' \emph{IEEE Transactions on Geoscience and Remote Sensing}, vol.~60, pp. 1--12, 2022.

\bibitem{wx22tgrs}
X.~Wu, D.~Hong, and J.~Chanussot, ``Convolutional neural networks for multimodal remote sensing data classification,'' \emph{IEEE Transactions on Geoscience and Remote Sensing}, vol.~60, pp. 1--10, 2022.

\bibitem{ldx24tgrs}
D.~Li, W.~Xie, Y.~Li, and L.~Fang, ``Fedfusion: Manifold-driven federated learning for multi-satellite and multi-modality fusion,'' \emph{IEEE Transactions on Geoscience and Remote Sensing}, vol.~62, pp. 1--13, 2024.

\bibitem{gyh23tgrs}
Y.~Gao, M.~Zhang, J.~Wang, and W.~Li, ``Cross-scale mixing attention for multisource remote sensing data fusion and classification,'' \emph{IEEE Transactions on Geoscience and Remote Sensing}, vol.~61, pp. 1--15, 2023.

\bibitem{zly23jstars}
L.~Zhao and S.~Ji, ``Cnn, rnn, or vit? an evaluation of different deep learning architectures for spatio-temporal representation of sentinel time series,'' \emph{IEEE Journal of Selected Topics in Applied Earth Observations and Remote Sensing}, vol.~16, pp. 44--56, 2023.

\bibitem{gyh22rs}
Y.~Gao, X.~Song, W.~Li, J.~Wang, J.~He, X.~Jiang, and Y.~Feng, ``Fusion classification of hsi and msi using a spatial-spectral vision transformer for wetland biodiversity estimation,'' \emph{Remote Sensing}, vol.~14, no.~4, pp. 1--19, 2022.

\bibitem{xzx22tgrs}
Z.~Xue, X.~Tan, X.~Yu, B.~Liu, A.~Yu, and P.~Zhang, ``Deep hierarchical vision transformer for hyperspectral and lidar data classification,'' \emph{IEEE Transactions on Image Processing}, vol.~31, pp. 3095--3110, 2022.

\bibitem{fyn24kbs}
Y.~Feng, J.~Zhu, R.~Song, and X.~Wang, ``S2eft: Spectral-spatial-elevation fusion transformer for hyperspectral image and lidar classification,'' \emph{Knowledge-Based Systems}, vol. 283, pp. 1--11, 2024.

\bibitem{zmq23rs}
M.~Zhang, F.~Gao, T.~Zhang, Y.~Gan, J.~Dong, and H.~Yu, ``Attention fusion of transformer-based and scale-based method for hyperspectral and lidar joint classification,'' \emph{Remote Sensing}, vol.~15, no.~3, pp. 1--15, 2023.

\bibitem{js23iclr}
J.~T. Smith, A.~Warrington, and S.~Linderman, ``Simplified state space layers for sequence modeling,'' in \emph{International Conference on Learning Representations}, 2023, pp. 1--13.

\bibitem{Mamba}
\BIBentryALTinterwordspacing
A.~Gu and T.~Dao, ``Mamba: Linear-time sequence modeling with selective state spaces,'' \emph{arXiv, 2312.00752}, 2024. [Online]. Available: \url{https://arxiv.org/abs/2312.00752}
\BIBentrySTDinterwordspacing

\bibitem{Mamba_tgrs1}
D.~Liao, Q.~Wang, T.~Lai, and H.~Huang, ``Joint classification of hyperspectral and lidar data base on mamba,'' \emph{IEEE Transactions on Geoscience and Remote Sensing}, pp. 1--15, 2024.

\bibitem{Mamba_jstar1}
Q.~Wang, X.~Fan, J.~Huang, S.~Li, and T.~Shen, ``Spectral--spatial feature extraction network with ssm--cnn for hyperspectral--multispectral image collaborative classification,'' \emph{IEEE Journal of Selected Topics in Applied Earth Observations and Remote Sensing}, vol.~17, pp. 17\,555--17\,566, 2024.

\bibitem{js24arxiv}
\BIBentryALTinterwordspacing
J.~Sieber, C.~A. Alonso, A.~Didier, M.~N. Zeilinger, and A.~Orvieto, ``Understanding the differences in foundation models: Attention, state space models, and recurrent neural networks,'' \emph{arXiv, 2405.15731}, 2024. [Online]. Available: \url{https://arxiv.org/abs/2405.15731}
\BIBentrySTDinterwordspacing

\bibitem{RSISCMDP}
G.~Cheng, X.~Xie, J.~Han, L.~Guo, and G.-S. Xia, ``Remote sensing image scene classification meets deep learning: Challenges, methods, benchmarks, and opportunities,'' \emph{IEEE Journal of Selected Topics in Applied Earth Observations and Remote Sensing}, vol.~13, pp. 3735--3756, 2020.

\bibitem{zgq23tgrs}
G.~Zhou, Y.~Tang, W.~Zhang, W.~Liu, Y.~Jiang, E.~Gao, Q.~Zhu, and Y.~Bai, ``Shadow detection on high-resolution digital orthophoto map using semantic matching,'' \emph{IEEE Transactions on Geoscience and Remote Sensing}, vol.~61, pp. 1--20, 2023.

\bibitem{zml23tbc}
M.~Zhou, L.~Chen, X.~Wei, X.~Liao, Q.~Mao, H.~Wang, H.~Pu, J.~Luo, T.~Xiang, and B.~Fang, ``Perception-oriented u-shaped transformer network for 360-degree no-reference image quality assessment,'' \emph{IEEE Transactions on Broadcasting}, vol.~69, no.~2, pp. 396--405, 2023.

\bibitem{cdq22tcsvt}
D.~Cheng, L.~Chen, C.~Lv, L.~Guo, and Q.~Kou, ``Light-guided and cross-fusion u-net for anti-illumination image super-resolution,'' \emph{IEEE Transactions on Circuits and Systems for Video Technology}, vol.~32, no.~12, pp. 8436--8449, 2022.

\bibitem{TBCNN}
X.~Xu, W.~Li, Q.~Ran, Q.~Du, L.~Gao, and B.~Zhang, ``Multisource remote sensing data classification based on convolutional neural network,'' \emph{IEEE Transactions on Geoscience and Remote Sensing}, vol.~56, no.~2, pp. 937--949, 2017.

\bibitem{PTop_CNN}
M.~Zhang, W.~Li, Q.~Du, L.~Gao, and B.~Zhang, ``Feature extraction for classification of hyperspectral and lidar data using patch-to-patch cnn,'' \emph{IEEE Transactions on Cybernetics}, vol.~50, no.~1, pp. 100--111, 2020.

\bibitem{FusAtNet}
S.~Mohla, S.~Pande, B.~Banerjee, and S.~Chaudhuri, ``Fusatnet: Dual attention based spectrospatial multimodal fusion network for hyperspectral and lidar classification,'' in \emph{2020 IEEE/CVF Conference on Computer Vision and Pattern Recognition Workshops (CVPRW)}, 2020, pp. 416--425.

\bibitem{S$^2$ENet}
S.~Fang, K.~Li, and Z.~Li, ``S$^2$enet: Spatial--spectral cross-modal enhancement network for classification of hyperspectral and lidar data,'' \emph{IEEE Geoscience and Remote Sensing Letters}, vol.~19, pp. 1--5, 2021.

\bibitem{hyx22igarss}
Y.~Hu, H.~He, and L.~Weng, ``Hyperspectral and {LiDAR} data land-use classification using parallel {Transformers},'' in \emph{IEEE International Geoscience and Remote Sensing Symposium (IGARSS)}, 2022, pp. 703--706.

\bibitem{HCT}
G.~Zhao, Q.~Ye, L.~Sun, Z.~Wu, C.~Pan, and B.~Jeon, ``Joint classification of hyperspectral and {LiDAR} data using a hierarchical cnn and {Transformer},'' \emph{IEEE Transactions on Geoscience and Remote Sensing}, vol.~61, pp. 1--16, 2022.

\bibitem{ExVit}
J.~Yao, B.~Zhang, C.~Li, D.~Hong, and J.~Chanussot, ``Extended vision transformer (exvit) for land use and land cover classification: A multimodal deep learning framework,'' \emph{IEEE Transactions on Geoscience and Remote Sensing}, vol.~61, pp. 1--15, 2023.

\bibitem{li2023mixing}
K.~Li, D.~Wang, X.~Wang, G.~Liu, Z.~Wu, and Q.~Wang, ``Mixing self-attention and convolution: A unified framework for multisource remote sensing data classification,'' \emph{IEEE Transactions on Geoscience and Remote Sensing}, vol.~61, pp. 1--16, 2023.

\bibitem{ga21nips}
A.~Gu, I.~Johnson, K.~Goel, K.~Saab, T.~Dao, A.~Rudra, and C.~Re, ``Combining recurrent, convolutional, and continuous-time models with linear state-space layers,'' in \emph{International Conference on Neural Information Processing Systems (NeurIPS)}, 2021, pp. 1--14.

\bibitem{ga22iclr}
A.~Gu, K.~Goel, and C.~R\'e, ``Efficiently modeling long sequences with structured state spaces,'' in \emph{International Conference on Learning Representations (ICLR)}, 2022, pp. 1--13.

\bibitem{ly24vmamba}
\BIBentryALTinterwordspacing
Y.~Liu, Y.~Tian, Y.~Zhao, H.~Yu, L.~Xie, Y.~Wang, Q.~Ye, and Y.~Liu, ``Vmamba: Visual state space model,'' \emph{arXiv, 2401.10166}, 2024. [Online]. Available: \url{https://arxiv.org/abs/2401.10166}
\BIBentrySTDinterwordspacing

\bibitem{rs3mamba}
X.~Ma, X.~Zhang, and M.-O. Pun, ``Rs3mamba: Visual state space model for remote sensing image semantic segmentation,'' \emph{IEEE Geoscience and Remote Sensing Letters}, vol.~21, pp. 1--5, 2024.

\bibitem{cky24rsmamba}
K.~Chen, B.~Chen, C.~Liu, W.~Li, Z.~Zou, and Z.~Shi, ``Rsmamba: Remote sensing image classification with state space model,'' \emph{IEEE Geoscience and Remote Sensing Letters}, vol.~21, pp. 1--5, 2024.

\bibitem{zqf24arxiv}
\BIBentryALTinterwordspacing
Q.~Zhu, Y.~Cai, Y.~Fang, Y.~Yang, C.~Chen, L.~Fan, and A.~Nguyen, ``Samba: Semantic segmentation of remotely sensed images with state space model,'' \emph{arXiv, 2404.01705}, 2024. [Online]. Available: \url{https://arxiv.org/abs/2404.01705}
\BIBentrySTDinterwordspacing

\bibitem{chen24cm}
H.~Chen, J.~Song, C.~Han, J.~Xia, and N.~Yokoya, ``Changemamba: Remote sensing change detection with spatiotemporal state space model,'' \emph{IEEE Transactions on Geoscience and Remote Sensing}, vol.~62, pp. 1--20, 2024.

\bibitem{lyp24tgrs}
Y.~Li, Y.~Luo, L.~Zhang, Z.~Wang, and B.~Du, ``Mambahsi: Spatial-spectral mamba for hyperspectral image classification,'' \emph{IEEE Transactions on Geoscience and Remote Sensing}, vol.~62, pp. 1--14, 2024.

\bibitem{Sigma}
Z.~Wan, P.~Zhang, Y.~Wang, S.~Yong, S.~Stepputtis, K.~Sycara, and Y.~Xie, ``Sigma: Siamese mamba network for multi-modal semantic segmentation,'' \emph{arXiv preprint arXiv:2404.04256}, 2024.

\bibitem{ESA}
``Multimodal remote sensing benchmark datasets for land cover classification with a shared and specific feature learning model,'' \emph{ISPRS Journal of Photogrammetry and Remote Sensing}, vol. 178, pp. 68--80, 2021.

\bibitem{berlin}
A.~Okujeni, S.~van~der Linden, and P.~Hostert, ``Berlin-urban-gradient dataset 2009-an enmap preparatory flight campaign,'' \emph{GFZ Data Services}, 2016.

\bibitem{ls18grsm}
B.~Le~Saux, N.~Yokoya, R.~Hansch, and S.~Prasad, ``2018 {IEEE GRSS} data fusion contest: Multimodal land use classification,'' \emph{IEEE Geoscience and Remote Sensing Magazine}, vol.~6, no.~1, pp. 52--54, 2018.

\bibitem{DFINet}
Y.~Gao, W.~Li, M.~Zhang, J.~Wang, W.~Sun, R.~Tao, and Q.~Du, ``Hyperspectral and multispectral classification for coastal wetland using depthwise feature interaction network,'' \emph{IEEE Transactions on Geoscience and Remote Sensing}, vol.~60, pp. 1--15, 2021.

\bibitem{AsyFFNet}
W.~Li, Y.~Gao, M.~Zhang, R.~Tao, and Q.~Du, ``Asymmetric feature fusion network for hyperspectral and sar image classification,'' \emph{IEEE Transactions on Neural Networks and Learning Systems}, vol.~34, no.~10, pp. 8057--8070, 2022.

\end{thebibliography}
\bibliographystyle{IEEEtran}

\end{document}